\documentclass[a4paper,10pt]{article}

\usepackage{amsfonts,amsmath,amsthm}
\usepackage{color,graphicx,float}
\usepackage{caption}
\usepackage{subcaption}
\usepackage{authblk}
\usepackage[UKenglish]{isodate}
\usepackage{amsopn}

\usepackage[top=3cm,left=2.5cm]{geometry}
\usepackage[italian,english]{babel} 
\newtheorem{theorem}{Theorem}[section]
\newtheorem{lemma}{Lemma}[section]

\newtheorem{definition}{Definition}[section]

\theoremstyle{remark}
\newtheorem{remark}{Remark}[section]
\newtheorem{example}{Example}[section]

\title{Landau-like theory for buckling phenomena and its application to the elastica hypoarealis}
\author{Stefano S.\,Turzi \thanks{\texttt{stefano.turzi@polimi.it}}}

\affil{\small{Dipartimento di Matematica, Politecnico di Milano, Piazza Leonardo da Vinci 32, 20133 Milano, Italy}}

\cleanlookdateon
\date{\today}

\newcommand{\Z}{\mathbb{Z}}
\newcommand{\R}{\mathbb{R}}
\newcommand{\Lop}{\mathbb{L}}

\newcommand{\bA}{\mathbf{A}}

\newcommand{\bN}{\mathbf{N}}

\newcommand{\bb}{\mathbf{b}}
\newcommand{\be}{\mathbf{e}}
\newcommand{\bff}{\mathbf{f}}
\newcommand{\bg}{\mathbf{g}}
\newcommand{\bn}{\mathbf{n}}
\newcommand{\br}{\mathbf{r}}
\newcommand{\bt}{\mathbf{t}}
\newcommand{\bv}{\mathbf{v}}
\newcommand{\bw}{\mathbf{w}}
\newcommand{\by}{\mathbf{y}}

\newcommand{\mn}[1]{\mathrm{m}_n^{#1}}

\renewcommand{\rho}{\varrho}
\newcommand{\eps}{\varepsilon}

\newcommand{\er}{\mathrm{e}}

\newcommand{\dr}{\mathrm{d}}

\newcommand{\D}[2]{\frac{\partial #1}{\partial #2}}

\newcommand{\Np}{N^{\perp}}
\newcommand{\cc}{\text{circ}}

\newcommand{\modif}[1]{{#1}}

\begin{document}

\maketitle 

\begin{abstract}
Bifurcation phenomena are ubiquitous in elasticity, but their study is often limited to linear perturbation or numerical analysis since second or higher variations are often beyond an analytic treatment. Here, we review two key mathematical ideas, namely, the splitting lemma and the determinacy of a function, and show how they can be fruitfully used to derive a reduced function, named Landau expansion in the paper, that allows us to give a simple but rigorous description of the bifurcation scenario, including the stability of the equilibrium solutions. We apply these ideas to a paradigmatic example with potential applications to various softly constrained physical systems and biological tissues: a stretchable elastic ring under pressure. We prove the existence of a tricritical point and find bistability effects and hysteresis when the stretching modulus is sufficiently small. These results seem to be in qualitative agreement with some recent experiments on heart cells.
\end{abstract}

%
%
%

\section{Introduction}
\label{sec:intro}

The traditional analysis of phase transitions in condensed matter physics involves the introduction of some key physical quantities: order parameters and Landau potential. An order parameter is a quantity which changes the value on going from one phase to the other and that can therefore be used to monitor the transition. It is often a non-trivial physical problem to find a suitable order parameter, and this can either be a scalar, vectorial or tensorial field. The phase, or state, of the system is then determined by minimizing a Landau potential, i.e, typically a polynomial function in the order parameter with coefficients depending on external control parameters. Again, finding a suitable Landau potential is often difficult and involves a number of simplifying steps. In physics this is a simplified theory, that neglects fluctuations, but explains the universality of phase transitions. Furthermore, symmetry aspects usually constitute a key part of the theory, and Landau's potential is precisely built to account for the changes in symmetry of the various phases. The main advantage of this approach is that it gives an intuitive account of phase changes: it is sufficient to study how the minima of a family of polynomials change as the external (or control) parameters are varied. It provides a complete scenario of all possible phase transitions and also yields information about the local stability of the critical points. As such, Landau theory can be seen as a basic tool to investigate bifurcation scenarios in microscopic theories, that would otherwise be nearly impossible to study given their intrinsic complexity.

The situation in elasticity is drastically different. Here, a well established elastic energy, whose minima determine the equilibrium configurations, is usually known in advance and does not need to be constructed from a microscopic energy. However, it is usually a parameter-dependent functional, rather than a function. Therefore, it is much more difficult to study since it lives in a infinite dimensional space. Furthermore, it is not clear in this context what the order parameters are and what they represent.

In this paper we show how it is possible to perform a rigorous finite dimensional reduction of the elastic energy to build a Landau-like potential (also called reduced energy function) to study buckling phenomena in elasticity and the local stability of the equilibrium solutions. 

The methods outlined in the paper are then applied to a problem in the mechanics of soft constrained materials. Namely, we study the bifurcations of an \emph{extensible} and closed elastic planar wire, when subject to an external pressure. \modif{Following Ref.\cite{02CapovillaEPJB}, we have used ``elastica hypoarealis'' in the title, from the suffix hypo (``less than normal''), and arealis (``pertaining to an area''), to describe the situation of a closed elastic rod which encloses an area that is smaller than the natural one.} The related problem of an inextensible elastic ring was stated and studied by Maurice L\'{e}vy in his memoir \cite{84Levy} more than a century ago, who investigated the equilibrium configurations of an infinitely long cylindrical pipe under pressure. This system is formally equivalent to a two-dimensional capillary film enclosed by a flexible rod, where the effects of pressure is substituted by the film surface tension, or to a bidimensional vesicle with an enclosed-area constraint \modif{\cite{13Giomi}}.

Despite that the analytic form of the equilibrium solutions is known (at least in the inextensible case) \cite{02CapovillaPRE,02CapovillaEPJB,13Giomi,08Vassilev}, this is difficult to grasp because it involves complicated elliptic functions. Furthermore, the nature of the bifurcations (or the order of the phase transitions, in physics parlance) is only determined via numerical experiments \cite{09Veerapaneni} or using heuristic arguments. For instance in \cite{09Veerapaneni} the stability of equilibrium shapes is investigated by adding a small perturbation and then performing a dynamical simulation with viscous damping. If the system relaxes back to the original equilibrium it is concluded that the shape is stable. 

This problem is considered as a paradigmatic example which is related to many problems in contemporary bio-physics such as the study of the equilibrium shapes of biological vesicles \cite{08Vassilev} or the mechanics of contractile cells adhering to soft substrates \cite{13Giomi}. For example, incorporating the bending elasticity of the actin cortex in a contour model of adherent cells \cite{13Banerjee,19Giomi}, seems to be essential to explain some experiments on cardiac myocytes \cite{11Chopra} and leads to an extremely rich polymorphism. In this case the cell boundary undergoes a discontinuous transition to a buckled configuration, accompanied by a region of bistability and hysteresis. This result is compatible with our theoretical prediction of a first-order transition from a circular to a buckled shape when the stretching modulus of the boundary is small. Other related problems include the dehydration dynamics of a hydrogel, where the enclosed water imposes a volume constraint \cite{18NardinocchiJAP,18NardinocchiSM} and shape transitions of carbon nanotubes under pressure \cite{04Zang}.

The paper is organized as follows. In \S\ref{sec:mathematical} we review the necessary mathematical background, which we illustrate with some examples in \S\ref{sec:EulerBeam}. The problem of a stretchable elastica under pressure is addressed in \S\ref{sec:variational}. In the following \S\ref{sec:linearanalysis} and \S\ref{sec:Landaupotential} we perform a complete splitting-lemma finite-dimensional reduction of the energy functional that allows us to study the transitions in \S\ref{sec:phasetransitions}. We draw the conclusions in \S\ref{sec:conclusions}. For ease of reading, the proof of the splitting lemma and some particularly long equations are reported in three appendices.

\section{Mathematical background}
\label{sec:mathematical}
Analysis of bifurcations is a vast area and many (closely connected) tools are already routinely used. For example: Lyapunov-Schmidt reduction, centre manifold theorem, Morse theory, elementary catastrophe theory. However, they are rarely used in elasticity to rigorously derive a Landau-like potential, starting from a stored elastic energy functional. We wish to show how tools from Singularity theory often allow us to reduce an infinite dimensional variational problem in elasticity to the analysis of the critical points of a \emph{family of polynomials in a finite number of variables}. We will refer to such a polynomial as the ``reduced function'' or the ``Landau expansion" of the original energy functional.

By contrast to the analysis of phase transitions in condensed matter physics, problems in elasticity theory generally have a natural variational formulation, where the final gaol is finding the minima of a given stored elastic energy $W$, as the material parameters are varied. 

Therefore, a typical mathematical setting is as follows \cite{84Geymonat,88Geymonat}. Let $E$ be a Banach space with scalar product $\langle,\rangle$; $U \subset E$ a neighborhood of $0 \in E$; $\Lambda$ a finite dimensional space (space of parameters); $W:E \times \Lambda \to \R$ a $C^k$ ($k \geq 3$) functional $W(f,\lambda)$. We want to study the number and the nature of the critical points of $W$ as the parameters $\lambda$ change. For simplicity of notation, we will sometimes leave the dependence on parameters $\lambda$ implicit. Furthermore, we assume that there exists a $C^{k-1}$-operator, $\nabla W: U \to E$, such that the Frechet derivative of $W$ is 
\begin{equation}
DW(f)[u] = \langle \nabla W(f), u \rangle
\end{equation}
for all $u \in E$. The equilibrium equations are written as $\nabla W(f)=0$.

\subsection{Splitting lemma}
It is known that a necessary condition for the presence of a bifurcation is that the hessian of $W$ is singular. It is only at those point that we cannot use the implicit function theorem and the number of critical points can change \cite{Ambrosetti}. Most of the times, however, the bifurcation involves only a finite and small number of variables. More precisely, in order to study the nature of the bifurcation it is often possible to perform a change of variable and transform the original functional into the sum of a regular part, where nothing really interesting happens, and a singular part, which develops a bifurcation. Remarkably, the singular part, also called \emph{reduced functional}, is often simply a function, so that the study of the bifurcation is reduced to a much simpler finite dimensional problem. This argument is made rigorous by the following Splitting lemma \cite{69Gromoll,75Chillingworth,83Golubitsky,84Geymonat}. For the reader's convenience we report the proof in Appendix \ref{app:proofSL}.

We assume that 
\begin{enumerate}
	\item $W(0)=0$ \label{thm:sl_ass1}
	\item $DW(0)[u] = 0$, for all $u \in U$ (0 is always an equilibrium solution) \label{thm:sl_ass2}
	\item $D^{2}W(0)[u,v] = \langle \Lop u, v \rangle$, where $\Lop$ is a symmetric Fredholm operator of index 0 \label{thm:sl_ass3}
\end{enumerate}

It is worth noticing that the equations $\Lop u = 0$ are the equilibrium equations, linearised about the trivial solution $u=0$. In fact,
\begin{align}
D^{2}W(0)[u,v] & = \lim_{\eps \to 0} \frac{1}{\eps}\big(DW(\eps v)[u] - DW(0)[u] \big) \notag \\
& = \lim_{\eps \to 0} \frac{1}{\eps}\big( \langle \nabla W(\eps v),u \rangle - \langle \nabla W(0),u \rangle \big)
 = \langle D\nabla W(0)[v],u \rangle,
\end{align}
so that $\Lop $ is the differential of $\nabla W$ at $u=0$
\begin{equation}
\Lop  = D(\nabla W(0))
\label{eq:L_and_gradW}
\end{equation}

Assumption \ref{thm:sl_ass3}. implies that (a) the ``Hessian operator'' $\Lop $ has \emph{finite-dimensional} and closed null-space $N$ and range; (b) $E$ is the decomposed as $E = N \oplus \Np$; (c) The restriction $\Lop  \big|_{\Np}$ is an isomorphism. These are the crucial assumptions that will allow us to reduce the problem to a finite dimensional one.

A point $x+y \in E$ with $x \in N$ and $y \in \Np$ is denoted as $(x,y) \in N\oplus \Np$, so that it is immediately clear which part of the sum belongs to $N$ and which part belongs to its orthogonal complement.

\begin{lemma}
Let $W$ be a functional as before. Then, there exist
\begin{enumerate}
\item an origin preserving $k$-diffeomorphism $\Phi:U \to E$ that fixes $x$, i.e., a change of coordinates of the form $(\bar x,\bar y) = \Phi(x,y) = (x, \eta(x,y))$, with $(0,0) = \Phi(0,0)$,
\item a differentiable map $h: U \cap N \to \Np: x \mapsto y=h(x)$, 
\end{enumerate}
such that locally $W$ splits into the sum of a regular and singular part
\begin{equation}
W \circ \Phi(x,y) = \underbrace{\frac{1}{2} \langle \Lop y, y \rangle}_{\text{regular part}} + \underbrace{W(x+h(x))}_{\text{singular part}}.
\end{equation}
Furthermore,
\begin{enumerate}
\setcounter{enumi}{2}
\item $y=h(x)$ is defined implicitly by the following equation. For a given $x \in U \cap N$, there exists a unique solution $y \in \Np$ of
\begin{equation}
Q \nabla W(x,y) = 0,
\label{eq:h}
\end{equation}
with $Q$ orthogonal projector onto $ \Np$ (with respect to $\langle, \rangle$);
\item $h(0)=0$, $h'(0)=0$;
\item Defining $g(x) =  W(x+h(x))$, $g(x)$ is a non-Morse function, i.e., $g(0)=0$, $g'(0)=0$, $g''(0)=0$.
\end{enumerate}
\label{thm:splitting}
\end{lemma}

\begin{figure}[H]
\begin{center}
\includegraphics[width=0.65\textwidth]{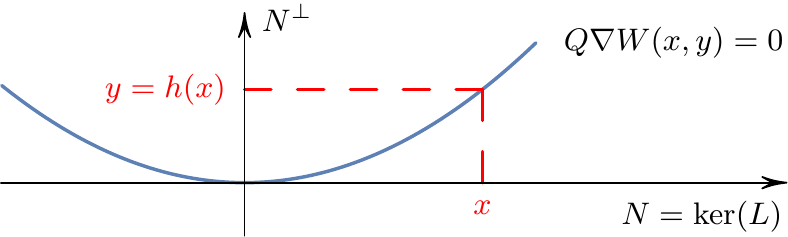} 
\end{center}
\caption{Schematic representation of the implicit definition of $y=h(x)$, as obtained by solving \eqref{eq:h}.}
\label{fig:schematic_h}
\end{figure}

\begin{remark}
\begin{enumerate}
\item The regular part, $\frac{1}{2} \langle \Lop y, y\rangle$, is not involved in the study of the bifurcations. In our case, it is just a positive quadratic form, because it is natural to assume that a physically sound elastic energy is bounded from below.  
\item The function $g(x)=W(x+h(x)): N \to \R$ is called the \emph{reduced function} and lives in the \emph{finite dimensional} space $N$. It captures all the qualitative information about the singularity and bifurcations.
\item Splitting lemma is the variational counterpart of the Lyapunov-Schmidt reduction \modif{\cite{MarsdenBook,03Golubitsky}} and splits the variables into ``essential'' ($x$) and ``inessential'' ($y$). The advantage over this method is that it readily provides information about the local stability of the critical points.
\item The diffeomorphism $\Phi$ is a change of variables that does not effect the number or the nature of singular points (but may affect their position or value).
\item Sometimes we say that the variable $y \in \Np$ is ``slaved'' to $x \in N$ in a neighborhood of 0. This is because, once $x$ is given, we can solve \eqref{eq:h} to uniquely define $y$: $y=h(x)$. This is illustrated in Fig.\ref{fig:schematic_h}. However, from the computational point of view, this is usually the most delicate step, that can be solved only using perturbation methods.
\end{enumerate}
\end{remark}

\subsection{Determinacy}
\label{sec:determinacy}

Most of the times we can only solve \eqref{eq:h} pertubatively, so that we are only able to calculate a Taylor expansion of $g(x)$. A key point, often overlooked, for a rigorous finite-dimensional reduction is the truncation of this power series. Is it possible to truncate the Taylor expansion at some suitable degree and still obtain a correct description of the bifurcation diagram? The answer is provided by the \emph{determinacy} of $g(x)$, defined below.

The theory deals with functions $g: U \to \R$ where $U$ is a neighborhood of 0. The cleanest way to handle such functions is to pass to germs, a germ being a class of functions which agree on suitable neighborhoods of 0. All operations on germs are defined by performing similar operations on representatives of their classes. In the following, \modif{we shall make no distinction between a germ and a representative function.}

\begin{definition}
Two functions $f$ and $g$ are \emph{right-equivalent} about $0$, and we write $f \sim_{r} g$, if there is an origin preserving change of variable (diffeomorphism) $\varphi$ such that $f=g \circ \varphi$.
\end{definition}	

Furthermore, we introduce the following standard definition: the $k$-jet of a \modif{function} $f$, written as $j^k(f)$, is the formal Taylor expansion of $f$ up to and including terms of degree $k$. 

\begin{definition}
A \modif{function} $g$ is \emph{$k$-determined} if, for any $f$ such that $j^k(g)=j^k(f)$, it follows that $g$ is right-equivalent to $f$.
\end{definition}	

The value of this definition lies in the fact that there are precise algebraic criteria for finding the determinacy of a function \cite{68Mather,83Golubitsky,Demazure,Gilmore}. Hence, this is a purely algebraic question, with an algorithmic solution. However, explicit computations can be difficult.

\begin{remark}
\begin{enumerate}
\item For what follows, it is important to observe that the determinacy of a one-variable \modif{function} is simply given by the first non-vanishing term in its Taylor expansion. 
\item However, in more than one dimension this is no longer the case. The \modif{function} $x^{5}+y^5$, for instance, is 6-determined (and not 5-determined), yet its Taylor expansion contains only fifth-degree terms. 
\item Of course, a $k$-determined \modif{function} is right-equivalent to its $k$-jet.
\item A \modif{function at a non-degenerate critical point, where the Hessian is non-singular,} is 2-determined. 
\end{enumerate}
\end{remark}	

Although not strictly necessary for the purposes of this paper, we briefly describe one algebraic criterion for the determinacy of a \modif{function} $g(x_1,\ldots,x_n)$. More details can be found in the references \cite{68Mather,83Golubitsky,Demazure,Gilmore}. Most results are based on the so-called {\em Jacobian ideal}, $J[g]$. This is the ideal generated by $\partial g/\partial x_{i}$, i.e., 
$J[g]  =  \{ h_{1}\partial g/\partial x_{1} + \dots + h_{n}\partial g/\partial x_{n} \}$  for arbitrary \modif{functions} $h_{i}$. Let $\mathrm{m}_n$ denote the set of \modif{functions} $g$ with $g(0)=0$. More generally, we denote with $\mn{k}$ the set of \modif{functions} in $\mathrm{m}_n$ such that all their partial derivatives of order less than $k$ vanish at 0. These powers of $\mathrm{m}_n$ form a descending chain, i.e., $\mathrm{m}_{n} \supseteq \mn{2} \supseteq \mn{3} \supseteq \dots$. It can be shown that $\mn{k}$ is an ideal (in the ring of infinitely differentiable \modif{functions} defined at $0$) generated by all monomials of homogeneous degree $k$. 
Finally, we have the following result, due to Mather \cite{Demazure}
\begin{theorem}
Let $g$ be a \modif{function}, and let $r>2$ be an integer. If
\begin{equation} 
\mn{r-1} \subset J[V] + \mn{r} ,
\label{eq:mather} 
\end{equation} 
then $g$ is $r$-determined.
\end{theorem}

\subsection{Illustrative examples}
\label{sec:EulerBeam}
The practical application of the mathematical tools described in \S\ref{sec:mathematical} is best illustrated with some key examples.

\begin{example} Of course, the splitting lemma also works in a finite dimensional setting. Consider, for example \cite{Castrigiano}, the function
\begin{equation}
W(x,y) = y^2 +2 y x^2 + x^2y^2.
\end{equation}
The gradient and the Hessian matrix are
\begin{equation}
\nabla W(x,y) = (2 x y^2+4 x y,2 x^2 y+2 x^2+2 y), \qquad
H(x,y) = 
\begin{pmatrix}
2 y^2+4 y & 4 y x+4 x \\
4 y x+4 x & 2 x^2+2
\end{pmatrix}.
\end{equation}
The point $(x,y) = (0,0)$ is a critical point ($\nabla W(0,0) =0$), so we can apply the Splitting lemma to separate $W$ into regular part and singular part, in a neighbourhood of $(0,0)$. The matrix $\Lop $ is $H(0,0)$, namely
\begin{equation}
\Lop  = \begin{pmatrix}
0 & 0 \\
0 & 2 
\end{pmatrix},
\end{equation}
whose null space $N$ is the $x$-axis: $N = \{(x,y) : y=0 \}$. The space $N$ is one dimensional with basis vector $\bv = (1,0)$. Hence, the projector $Q$ is
\begin{equation}
Q = \begin{pmatrix}
0 & 0 \\
0 & 1 
\end{pmatrix}.
\end{equation}
The variable $y$ is slaved to $x$ and the map $y=h(x)$ is calculated by solving the equation $Q\nabla W(x,y) = 0$ with $y \in \Np$, for a given $x \in N$. Specifically, we get
\begin{align}
Q\nabla W(x,y) 
= \begin{pmatrix}
0 & 0 \\
0 & 1 
\end{pmatrix}
\begin{pmatrix}
2 x y^2+4 x y \\
2 x^2 y+2 x^2+2 y
\end{pmatrix}
 =
\begin{pmatrix}
0\\
2 x^2 y+2 x^2+2 y
\end{pmatrix}
=
\begin{pmatrix}
0\\
0
\end{pmatrix}.
\end{align}
In this simple example, the equation $x^2 y+x^2+y = 0$ can be solved explicitly for $y$ to give
\begin{equation}
y = h(x) = -\frac{x^2}{1+x^2}.
\end{equation}
After \modif{some algebra}, we can write the reduced functional as
\begin{equation}
g(x) = W(x,h(x)) = -\frac{x^4}{1+x^2}.
\end{equation}
To find the change of variables $\Phi:(x,y) \mapsto (x,\eta(x,y))$, we use the fact that 
\begin{equation}
W(x,\eta(x,y))-g(x)
\end{equation}
is a Morse function of the form $\tfrac{1}{2} \langle \Lop y, y \rangle = y^2$. Therefore,
\begin{align}
W(x,\eta(x,y)) + \frac{x^4}{1+x^2} = \eta^2 +2 \eta x^2 + x^2 \eta^2 + \frac{x^4}{1+x^2}
= y^2
\end{align}
This yields the diffeomorphism
\begin{equation}
\Phi(x,y) = \big(x, \eta(x,y) \big) = \left(x, -\frac{x^2}{1+x^2} + \frac{y}{\sqrt{1+x^2}} \right).
\end{equation}
With this substitution we find
\begin{equation}
W\circ \Phi(x,y) = y^2 - \frac{x^4}{x^2+1}.
\end{equation}
Most of the times, however, we don't need to find the change of variables $\Phi$ explicitly. It suffices to find the reduced function $g(x)$, and the most difficult step is the calculation of $y=h(x)$.
\end{example}

\begin{example}  Next, we take a classical example of nonlinear bifurcation in elasticity: the problem of the elastica. Analogous bifurcations are ubiquitous in physical problems, take for example the classical Fr{\'e}edericksz transition in liquid crystal physics \cite{deGennes,12Bevilacqua}. 
	
Consider the deformation of a slender inextensible elastic rod, subjected to a pair of compressive axial forces $F$ at both ends. By the Bernoulli-Euler beam theory, the bending moment is proportional to the curvature. When $F$ is less than a critical value, the beam remains straight in a horizontal position. When the critical load is exceeded, the beam buckles in a new state, since the straight configuration becomes unstable. Standard arguments in elasticity leads to the energy functional \modif{\cite{10Audoly}}
\begin{equation}
W(\theta) = \int_{0}^{L}\left(\frac{k}{2}(\theta')^2 + F \cos\theta\right)\dr s - F L,
\label{eq:WEuler}
\end{equation}	
where $L$ is the rod length, $s$ is the arc-length, $k$ is the bending modulus, $F$ is the axial compressive force and $\theta(s)$ is the inclination angle with respect to the $x$-axis. The constant term $FL$ has no physical effect, but it has been chosen for mathematical convenience, so that $W(0)=0$. We consider the case with hinge supports at two ends, so that the boundary conditions are $\theta'(0)=0$ and $\theta'(L)=0$. From a physical standpoint these correspond to a vanishing bending moment at both ends. Furthermore, we take $E=H^1(0,L)$ and $\langle,\rangle$ as the standard scalar product in $L^2$.\\
The equilibrium equation (Euler-Lagrange equation of \eqref{eq:WEuler}) is found to be
\begin{equation}
\nabla W (\theta) := -k\theta''-F \sin\theta = 0.
\label{eq:DW_Euler}
\end{equation}
It is easy to check that $\theta = 0$ (i.e., the undeformed configuration) is an equilibrium solution for any value of $F$. The linearization of \eqref{eq:DW_Euler} around the trivial solution $\theta=0$ yields the operator $\Lop $
\begin{equation}
\Lop u := -k u''-F u.
\label{eq:L_Euler}
\end{equation}
The linear problem, with $u'(0)=u'(L)=0$, has only the trivial solution (and thus $\ker(\Lop ) = \{0\}$) unless $F$ has one of the critical values
\begin{equation}
F/k=(n \pi/L)^2, \qquad n = 1,2,\ldots .
\end{equation}
For simplicity, we only consider the first buckling mode and take $n=1$. When $F = F_{cr} = k (\pi/L)^2$, there exists a non-trivial solution to $\Lop u=0$. In such a case, the null-space $N=\ker(\Lop )$ is one-dimensional and is generated by
\begin{equation}
v_0(s) =\cos(\pi s/L).
\end{equation}
The projector $Q:E \to \Np$ is
\begin{equation}
Qf = f - \langle f,v_0\rangle\, \frac{v_0}{\| v_0 \|^2} =  f(s) - \frac{2}{L}\left[\int_{0}^{L} f(s)v_0(s)\,\dr s \right] v_0(s), 
\end{equation}
with $\| v_0 \|^2 = L/2$. The solution is then written as $\theta(s) = \alpha\, v_0(s) + w(\alpha,s)$, where $\alpha$ is an \emph{arbitrary constant} ($v_0$ is an eigenvector of $\Lop $, with null eigenvalue, and its amplitude is arbitrary), and $w \in \Np$ is orthogonal to $v_0$ (in other words, $w$ has no $\cos(\pi\,s/L)$ term in its Fourier expansion). Furthermore, $w$ is determined, once we fix $\alpha$, by solving \eqref{eq:h}, namely $Q\nabla W(\alpha\,v_0 + w)=0$, with respect to $w$.

The constant $\alpha$ is what plays the role of an order parameter in a Landau-like theory of buckling: it vanishes in one ``phase'' (the straight configuration), while it is $\alpha \neq 0$ in the buckled ``phase''. Therefore it is a quantity which changes the value on going from one phase to the other and that can therefore be used to monitor the transition.

The main computational challenge is how to find $y=h(x)$. Specifically, for a given $\alpha$, we look for solutions of \eqref{eq:h} of the form $\theta(s) = \alpha\, v_0(s) + w(\alpha,s)$. Namely, the equation for $w(\alpha,s) \in \Np$ is
\begin{align}
& k \big(\alpha\, v_0'' + w''\big) + F_{cr} \sin(\alpha\, v_0 + w) \notag \\
& -\frac{2}{L} \cos (\pi s/L) \int_0^L \cos (\pi s/L) \left[k \big(\alpha\, v_0'' + w''\big) + F_{cr} \sin(\alpha\, v_0 + w)\right] \dr s = 0 ,
\label{eq:QgradW}
\end{align}
where $(\cdot)'$ denotes differentiation with respect to $s$. However, we can only solve this equation perturbatively, so that we substitute
\begin{equation}
w(\alpha,s) = \alpha^2 w_2(s) + \alpha^3 w_3(s) + O(\alpha^4).
\label{eq:w_expansion}
\end{equation}
and Taylor expand with respect to $\alpha$. By collecting the terms of homogeneous degree in $\alpha$, we obtain a chain of linear differential equations that it is possible to solve. In so doing, the Taylor expansion of $h(x)$, and then that of $g(x)$, is constructed. It is important to remark that \eqref{eq:w_expansion} does not contain zeroth and first degree terms, because the Splitting lemma guarantees that $h(0)=0$ and $h'(0)=0$ (i.e., $w(0,s)=0$ and $w_{\alpha}(0,s)=0$).
The substitution of \eqref{eq:w_expansion} into \eqref{eq:QgradW} yields, after some algebra,
\begin{align}
w_2'' + \frac{\pi^2}{L^2} w_2 = 0 , \qquad 
w_3'' + \frac{\pi^2}{L^2} w_3 = \frac{\pi^2}{24 L^2} \cos(3\pi s/L),
\end{align}
whose solutions, with $w_i'(0) = 0$, $w_i'(L) = 0$, and considering that $w_i \in \Np$ (no $\cos(\pi s /L)$ terms), are
\begin{align}
w_2(s)=0, \qquad w_3(s) = -\frac{1}{192}\cos (3\pi s/L).
\end{align}
Therefore, the sum that we have denoted with $x+h(x)$ in the Splitting lemma is, in our case
\begin{equation}
\theta(s) = \alpha \cos(\pi\,s/L) - \frac{\alpha^3}{192}\cos (3\pi s/L) + O(\alpha^4).
\label{eq:x+h_Euler}
\end{equation}
To calculate the reduced energy function $g(\alpha)$, we substitute \eqref{eq:x+h_Euler} back in $W$, Taylor expand the integrand with respect to $\alpha$, and calculate the integral term by term. The reduced energy function $g(\alpha) = W(\alpha \,v_0(s) + w(\alpha,s))$ contains all the necessary qualitative information about the bifurcations and the stability of solutions. \modif{At variance with weakly nonlinear analysis, the material parameters are not perturbed and the amplitude of the perturbation is not determined from a bifurcation equation. This is found by looking at the critical points of the reduced functional, where the original dependence on the material parameters is retained.}

After some calculations, not reported for brevity, this is found to be
\begin{align}
g(\alpha) = \left(\frac{\pi ^2 k}{4 L} - \frac{F L}{4}\right)  \alpha^2 + \frac{FL}{64}\alpha^4 + O(\alpha^6),
\label{eq:g_Euler}
\end{align}
which corresponds to a Landau-like potential and shows that the bifurcation is a classic supercritical pitchfork (second order transition). The local stability is also easily determined by looking at the minima/maxima of $g(\alpha)$ as the parameters vary.\\

However, two issues remain to be discussed: (1) is it sufficient to include in \eqref{eq:x+h_Euler} only $O(\alpha^k)$ terms, with $k\leq 3$? (2) is it appropriate to truncate the reduced potential $g(\alpha)$ at $O(\alpha^4)$? Both of these questions can be addressed by studying the determinacy of $g(\alpha)$, under maximum degeneracy conditions. Since $g(\alpha)$ is single-variable, its determinacy is given by the degree of the first non-vanishing term. We see from \eqref{eq:g_Euler} that it is possible to choose $F$, $k$ and $L$ such that the second degree term is zero. This condition yield the critical value $F_{cr}=k (\pi/L)^2$, and a corresponding non-Morse potential. However, with this choice of the parameters, the fourth degree term is necessarily different from zero, so that $g(\alpha)$ is at most 4-determined. This means that it is right-equivalent to its fourth-degree truncated Taylor expansion for any choice of the parameters
\begin{equation}
g(\alpha) \sim_{r} \left(\frac{\pi ^2 k}{4 L} - \frac{F L}{4}\right)  \alpha^2 + \frac{FL}{64}\alpha^4,
\label{eq:g4_Euler}
\end{equation}
and we only need to keep the terms in \eqref{eq:w_expansion} and \eqref{eq:x+h_Euler} that have a non-vanishing contribution to $g(\alpha)$, up to the fourth degree. Interestingly, it turns out that $w_3$ is necessary only to calculate the sixth-degree or higher terms, so that only the linear term, $\alpha v_0(s)$, in \eqref{eq:x+h_Euler} contributes to the calculation of \eqref{eq:g4_Euler}. In other words, the simple Euler beam problem can be successfully analysed just by linearizing the equilibrium equation, and using the linearized solution \cite{08Napoli}. This is exactly what is done in most Engineering literature. 

As a final remark, we note that this self-consistency procedure can only be performed \emph{a-posteriori}, once the expansion \eqref{eq:w_expansion} and the corresponding reduced function \eqref{eq:g_Euler} are determined with a sufficient number of terms, up to the maximum determinacy of $g(\alpha)$.
\end{example}

\modif{
\begin{example}
Finally, let us briefly sketch the finite-dimensional reduction of an \emph{extensible} elastic rod, subjected to a compressive axial force $F$ and hinges at both ends. The calculations are similar to the previous example, but the phase transition can either be first or second order depending on the stretching modulus $b$ of the rod \cite{01magn,15natu,17natu}. The geometry of the rod is now described by two variables, namely, by the angle $\theta(S)$ and the stretch $\lambda(S)$, where $S$ is the \emph{referential} arc-length and, by definition, $\lambda(S) = ds/dS$. The elastic energy is 
\begin{equation}
W(\theta,\lambda) = \int_{0}^{L}\left(\frac{k}{2}(\theta')^2 + \frac{b}{2}(\lambda-1)^2 
+ F \lambda\cos\theta\right)\dr S,
\label{eq:WEulerCompressible}
\end{equation}	
where the second term represents stretching energy and we have omitted the (physically unimportant) constant terms for simplicity. The equilibrium equations are
\begin{align}
k\theta'' + F \lambda \sin\theta = 0, \\
b(\lambda-1) + F \cos\theta = 0.
\label{eq:DW_EC2}
\end{align}
The second equation can be used to eliminate $\lambda(S)$ so that the energy \eqref{eq:WEulerCompressible} reads \cite{01magn}
\begin{equation}
W_{s}(\theta) = \int_{0}^{L}\left(\frac{k}{2}(\theta')^2 + F \cos\theta - \frac{F^2}{2b} (\cos\theta)^2 \right)\dr S.
\label{eq:WEulerC2}
\end{equation}	
The corresponding equilibrium equation for $\theta$ is
\begin{align}
k\theta'' + F \sin\theta -\frac{F^2}{2b} \sin(2\theta) = 0,
\end{align}
and it is immediate to check that $\theta = 0$ is always a solution. The linearisation of this equation (with boundary conditions $\theta'(0)=\theta'(L)=0$) yields the bifurcation condition
\begin{equation}
\frac{F}{k}\left(1-\frac{F}{b} \right) = (n \pi/L)^2.
\label{eq:criticalF_compressible}
\end{equation}
The analysis of this equation shows that mode $n$ cannot bifurcate when $bL^2 < 4k n^2 \pi^2$. As described in \cite {01magn}, the inextensible limit $bL^2/k \to +\infty$ implies that infinitely many buckling loads exist, whereas for the extensible case only a limited number of buckling loads, compatible with \eqref{eq:criticalF_compressible}, exist. For concreteness, we take $n=1$ and $bL^2 > 4k \pi^2$ and define the dimensionless ratios
\newcommand{\muhu}{\hat{\mu}_1}
\newcommand{\muhd}{\hat{\mu}_2}
\begin{equation}
\muhu = \frac{F}{b}, \qquad \muhd = \frac{bL^2}{\pi^2 k},
\end{equation}
which represent the compressive force measured in units of $b$ and the stretching to bending ratio\footnote{This is related to the square of the slenderness.}. It is worth noticing that the incompressible limit corresponds to $\muhd \to + \infty$, $\muhu \to 0$ such that $\muhu \muhd$ is constant. In such a case, in fact, we have that both $F \ll b$ and $k/L^2 \ll b$, but $F$ and $k/L^2$ are of the same order $FL^2/k = const.$ which reproduces the classical Euler elastica.

The bifurcation condition \eqref{eq:criticalF_compressible}, with $n=1$, reads $\muhu \muhd (1-\muhu) = 1$. The null space $N$ is again generated by $v_0(S) = \cos(\pi S/L)$, but now the asymptotic solution of Eq. \eqref{eq:h} yields
\begin{align}
\theta(S) & = \alpha  \cos (\pi  S/L) 
- \alpha^3 \frac{1 - 4\muhu}{192 (1-\muhu)} \cos (3 \pi  S/L)
+ O(\alpha^5).
\end{align}
Hence, the corresponding reduced functional, neglecting the constant terms, is found to be
\begin{align}
\frac{L}{\pi^2 k} g(\alpha) & = \frac{1}{4}(1-\muhu\muhd+\muhu^2\muhd) \alpha^2
+ \frac{1}{64}\muhu \muhd (1-4 \muhu) \alpha ^4 \notag \\
& + \frac{144 \muhu^4 \muhd-280 \muhu^3 \muhd+\muhu^2 (145 \muhd+16)-\muhu (9 \muhd+8)+1}{16384 (\muhu-1)^2} \, \alpha ^6,
\label{eq:g_EuleroComprimibile}
\end{align}
and it is possible to show that the expansion of $\theta(S)$ to $O(\alpha^3)$ is sufficient to reconstruct the energy coefficients exactly, up to sixth degree. The $O(2)$ and $O(4)$ coefficients, $a_2(\muhu,\muhd)$ and $a_4(\muhu,\muhd)$, both vanish in $(\muhu,\muhd) = (1/4,16/3)$ which then corresponds to the point of maximum degeneracy. At this point, however, the sixth-degree coefficients is positive, so that $g$ is 6-determined. Therefore, the transition can either be first ($a_4 <0$) or second order ($a_4 >0$).
\end{example}	
}

\section{\modif{Extensible elastic ring under uniform pressure}}
\label{sec:variational}
\begin{figure}
\begin{center}
\includegraphics[width=0.5\textwidth]{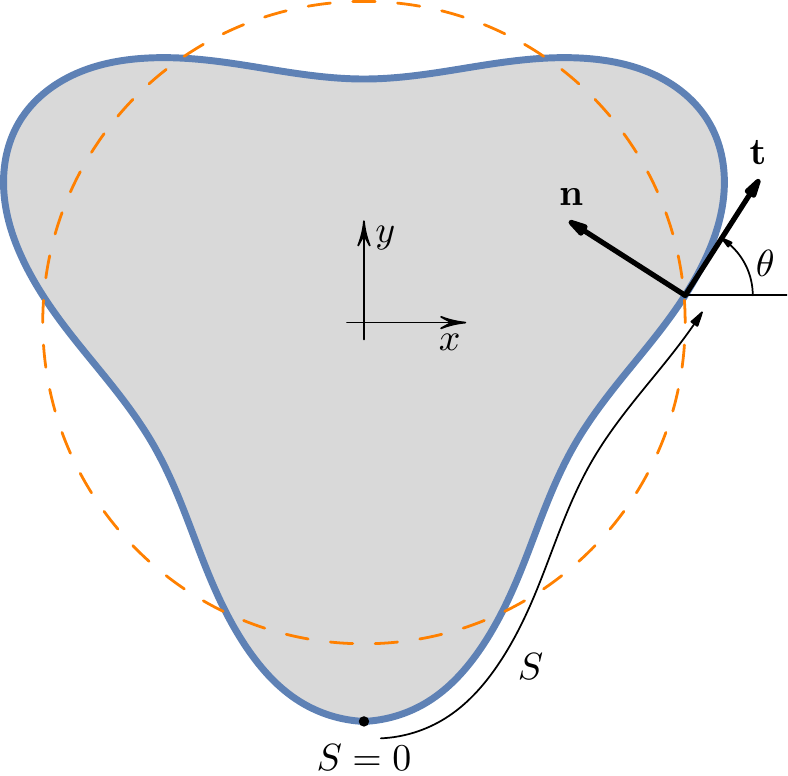}
\end{center}
\caption{Schematic representation of the shape profile. For comparison it is shown also the circular solution (dashed line).}
\label{fig:shape_riferimento}
\end{figure}

We assume a translational invariance in the longitudinal direction so that all the forces act on the cross-section of our tridimensional system. Hence, we can restrict to an equivalent two-dimensional system composed by flexible and \emph{stretchable} closed elastic rod, subject to a pressure difference (see Fig.\ref{fig:shape_riferimento}). The compressive normal force can be physically realised by an external pressure, a surface tension generated by, for example, a soap-film bounded by the elastic rod, or an area constraint induced by the confinement of an incompressible fluid.

\subsection{Description of the variational model}
\label{sec:model}
In a reference frame $(O;x,y)$ with coordinate unit vectors $\be_x$ and $\be_y$, the rod profile is modelled as a closed parametric curve $\br(S) = (x(S),y(S))$, with $S \in [0, L]$, where $L$ is the rod-length in a stress-free configuration and $S$ is the \emph{referential} arclength. Let $D$ be the region enclosed by the rod. We denote with $\theta(S)$ the inclination angle of the rod with respect to the $x$-axis. The unit tangent and unit normal are written as $\bt = \cos\theta \,\be_x + \sin\theta \,\be_y$ and $\bn = -\sin\theta \,\be_x + \cos\theta \,\be_y$. If we use $s$ for the arc-length in the deformed configuration, we readily obtain the identities
\begin{align}
\br'(S) = \frac{\dr \br}{\dr S} = \frac{\dr \br}{\dr s}\frac{\dr s}{\dr S} = \lambda(S)\,\bt(S),
\end{align}
where $\lambda(S) = \dr s/\dr S$ is the stretch ratio. The functional representing the stored elastic energy is
\begin{align}
W_{\text{el}}(\br,\theta,\lambda) &= \int_{0}^{L} \Big(\frac{k}{2}\theta'(S)^2 + \frac{b}{2}(\lambda(S)-1)^2 \Big)\,\dr S \notag \\
& -\int_{0}^{L} \bN(S) \cdot \big( \br'(S) - \lambda(S) \bt(S) \big)\,\dr S 
 + p \int_D \dr a
\label{eq:Wg}
\end{align}
where $k$ and $b$ are the bending and stretching moduli, $\dr a$ is the area element, $p$ is a surface tension or a hydrostatic pressure. The first integral comprises the bending energy and the stretching energy. The second integral enforces the identity $\br'=\lambda\bt$ with a Lagrange multiplier $\bN(S)$ that is interpreted as internal stress. In so doing, the energy is a functional of three \emph{independent} quantities, namely, $\br$, $\theta$ and $\lambda$. The last integral can be calculated as a line integral over the boundary of $D$:
\begin{equation}
\int_D \dr a = \frac{1}{2}\int_{0}^{L} \big(x(S) y'(S)-y(S) x'(S)\big)\dr S . 
\label{eq:area}
\end{equation}

\subsection{Equilibrium equations and reduced functional}
\label{sec:equilibrium_equations}

The equilibrium profile is governed by the Euler-Lagrange equations of \eqref{eq:Wg}. After some algebra, we find \cite{GorielyBook}
\begin{align}
k \theta'' - \lambda \bN\cdot \bn & = 0, \label{eq:eq1}\\
\big(N_x + p y \big)' & = 0 \label{eq:eq2}\\
\big(N_y - p x \big)' & = 0 \label{eq:eq3}\\
b(\lambda - 1) + \bN\cdot\bt & = 0, \label{eq:eq4}
\end{align}
where $\bN = N_x \be_x + N_y \be_y$, and we recall that $\bt = \cos\theta \,\be_x + \sin\theta \,\be_y$, $\bn = -\sin\theta \,\be_x + \cos\theta \,\be_y$. It is clear from \eqref{eq:eq2},\eqref{eq:eq3} that the internal stress has the form $N_x(S) = -py(S) + C_x$, $N_y(S) = px(S) + C_y$, where $C_x$ and $C_y$ are integration constants that can depend on the pressure, but, of course, cannot depend on $S$ and thus cannot depend on the shape profile $x(S)$ and $y(S)$.
Furthermore, from \eqref{eq:eq4} we can immediately find $\lambda(S)$
\begin{equation}
\lambda(S) = 1 - \frac{\bN\cdot\bt}{b} = 1 - \frac{N_x\cos\theta + N_y\sin\theta}{b},
\label{eq:lambda}
\end{equation}
and substitute its value in the energy, so to reduce the number of unknowns.

The circular configuration, where the material is simply compressed, has radius and area given by
\begin{equation}
R_{\cc} = \Big(\frac{2 \pi }{L}+\frac{p}{b} \Big)^{-1}, \qquad A_{\cc} = \pi R_{\cc}^2,
\end{equation}
and
\begin{subequations}
\begin{align}
\theta_{\cc}(S) = 2\pi S/L, \qquad  \lambda_{\cc} = (L/2\pi - R_{\cc})/(L/2\pi), \\
x_{\cc}(S) = R_{\cc} \sin (2\pi S/L), \qquad 
y_{\cc}(S) = -R_{\cc} \cos (2\pi S/L).
\end{align}
\label{eq:circular}
\end{subequations}
We recall that $L$ is a referential length and, thus, it does not represent the actual length of the rod, when deformed. The substitution of \eqref{eq:circular} into Eqs.\eqref{eq:eq1}-\eqref{eq:eq4}, i.e., the requirement that the circular solution is always an equilibrium solution, albeit possibly unstable, for any value of $p$, yields $C_x=C_y=0$, $(\bN\cdot\bn)_{\cc}=0$, $(\bN\cdot\bt)_{\cc} = p R_{\cc}$. Since the constants are independent of the particular configuration, we take
\begin{align}
N_x = p y, \qquad N_y = - p x,
\label{eq:N}
\end{align}
in general, and substitute \eqref{eq:N}, \eqref{eq:lambda} into \eqref{eq:Wg} in order to simplify the elastic energy and obtain a reduced functional in the variables $(\br, \theta)$:
\begin{align}
W_{r}(\br,\theta) &= \int_{0}^{L} \Big[\frac{k}{2} (\theta')^2 - p \big(y \cos\theta - x \sin\theta\big) \notag \\
& - \frac{p^2}{2b} \big(y\cos\theta - x\sin\theta\big)^2
- \frac{p}{2} \left(x y' - y x'\right) \Big] \dr S.
\label{eq:Wg_reduced}
\end{align}
It is possible to show that the Euler-Lagrange equations obtained form \eqref{eq:Wg_reduced} coincide with the equilibrium equations, when these are simplified by means of \eqref{eq:N}, \eqref{eq:lambda}. The reduced energy \eqref{eq:Wg_reduced} can be explicitly evaluated in the circular solution \eqref{eq:circular} and is equal to
\begin{align}
W_{\cc} = W_{r}(\br_{\cc},\theta_{\cc}) = 2\pi^2 \frac{k}{L} + p R_{\cc}(L-\pi R_{\cc}) - \frac{p^2}{2b}L R_{\cc}^2.
\label{eq:Wg_r_circular}
\end{align}

It is convenient to perform a point transformation of \eqref{eq:Wg_reduced} and change the original variables $(x,y)$ into $(\xi,\eta)$, defined as
\begin{equation}\begin{cases}
\xi = x \cos\theta + y \sin\theta, \\
\eta = -x \sin\theta + y \cos\theta,
\end{cases}
\label{eq:change_xieta}
\end{equation}
which corresponds to a simple rotation of the original variables $(x,y)$, and, as such, can be easily inverted once the angle $\theta$ is known. With this transformation, the reduced functional simplifies to
\begin{align}
W_{r}(\xi,\eta,\theta) & = \int_{0}^{L} \Big[\frac{k}{2} (\theta')^2 
- p \eta - \frac{p^2}{2 b} \eta^2 - \frac{p}{2} \left(\xi \eta' - \eta \xi'\right)
- \frac{p}{2} \left(\eta^2 + \xi^2 \right) \theta' \Big] \dr S,
\label{eq:Wg_reduced2}
\end{align}
and the circular solution is simply written as $\xi_{\cc}=0$, $\eta_{\cc} = -R_{\cc}$. 

Finally, we rescale the functional $W_{r}$ so that it can be studied using a Splitting lemma method. Namely, we consider the functional
\begin{equation}
W(\xi.\eta,\theta) = W_{r}(\xi,\eta,\theta) - W_{\cc},
\label{eq:W_final}
\end{equation}
and look for a finite dimensional reduction of $W$ in a neighbourhood of the circular solution.

The equilibrium equations (i.e., $\nabla W = 0$) read
\begin{subequations}
\begin{align}
k \theta'' - p \big(\xi \xi' + \eta \eta' \big) & = 0, \label{eq:eq1_b} \\[2mm]
\eta' + \theta' \xi & = 0,  \label{eq:eq2_b} \\
\xi' - \Big(\frac{p}{b} + \theta' \Big)\eta & = 1. \label{eq:eq3_b}
\end{align}
\label{eq:eq_hypoarealis}
\end{subequations}
Once the profile is known, the enclosed area as a function of $p$ is calculated from \modif{\eqref{eq:area}. In terms of the functions $\xi$, $\eta$ and $\theta$, this rewrites as}
\begin{equation}
A = \frac{1}{2}\int_{0}^{L} \big[\xi (\xi \theta' + \eta') + \eta (\eta \theta' - \xi')\big] \dr S.
\label{eq:Area_xieta}
\end{equation}

\section{Linear analysis and bifurcation condition}
\label{sec:linearanalysis}
The compressed circular shape is the unique equilibrium solution when $p$ is close to zero, but non-trivial solutions bifurcate off the circular solution when $p$ is increased. The critical threshold is obtained by perturbing the compressed circular solution and taking the linearised equilibrium equations. 

\modif{Clearly, the energy is unaffected by rigid translations. Therefore, we select one representative among all possible shifted solutions by taking the centre of the referential disk at the origin $O$. Furthermore, the energy \eqref{eq:W_final} is also fixed by the action of the rotation group $SO(2)$, and this is physically plausible, since any finite rotation of the solution yields an energetically equivalent configuration. We choose a representative solution with horizontal tangent at $S=0$ (i.e., we choose $\theta(0)=0$).} This, however, does not completely remove the degeneracy, since there is still the stabilizer subgroup $\Z_4$, the cyclic group which relabels the axis and is generated by the action 
\begin{align}
(\theta, x, y) \longmapsto (\theta - \tfrac{\pi}{2}, y, -x).
\end{align}
We could take advantage of this symmetry and use an equivariant formulation of the splitting lemma \cite{15Chillingworth}. However, this is not strictly necessary in our case, and introduces additional complications, so we prefer here to use the standard splitting lemma, \modif{and select representative solutions a-posteriori, by fixing its centre and a horizontal tangent in $S=0$.}

It is convenient to introduce the dimensionless parameters
\begin{align}
\mu_1 = \frac{pL}{2\pi\,b}, \qquad \mu_2 = \frac{b \, L^2}{k}
\end{align}
where $\mu_1$ measures the strength of surface tension relative to stretching and $\mu_2$ is the stretching to bending ratio. It is natural to assume  $\mu_2 \in (0,+\infty)$ with $\mu_2=0$ being the ``infinitely soft'' limit, and $\mu_2 \to +\infty$ the inextensible limit. 

\begin{figure}
\begin{tabular}{c}
\includegraphics[width=0.95\textwidth]{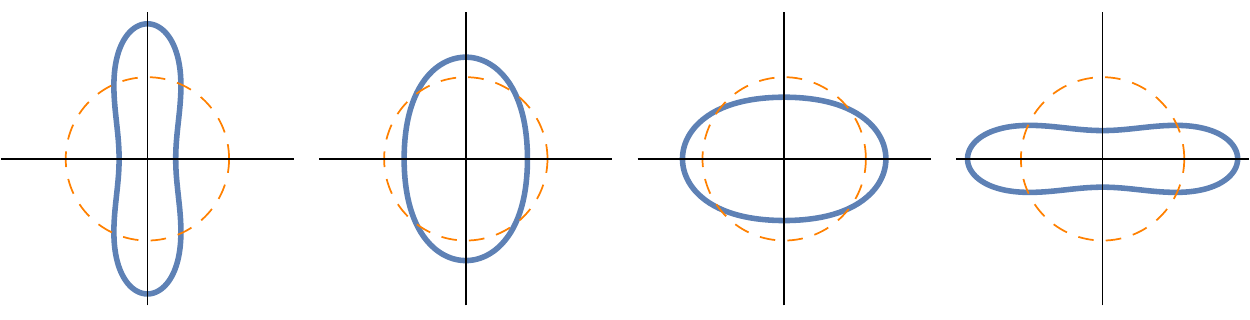} \\
\qquad $\alpha = -0.08$ \hfill $\alpha = -0.03$ \hfill $\alpha = 0.03$ \hfill $\alpha = 0.08$ \qquad\qquad \\[2mm]
\includegraphics[width=0.95\textwidth]{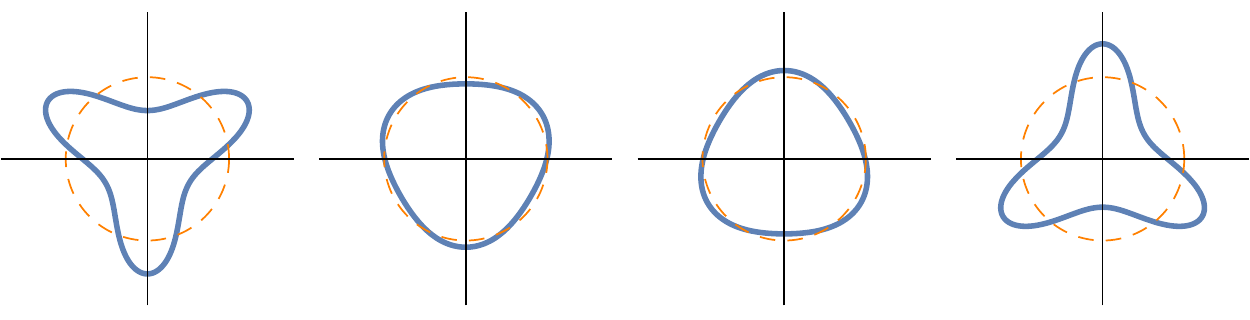} \\
\qquad $\alpha = -0.05$ \hfill $\alpha = -0.01$ \hfill $\alpha = 0.01$ \hfill $\alpha = 0.05$ \qquad\qquad \\[2mm]
\includegraphics[width=0.95\textwidth]{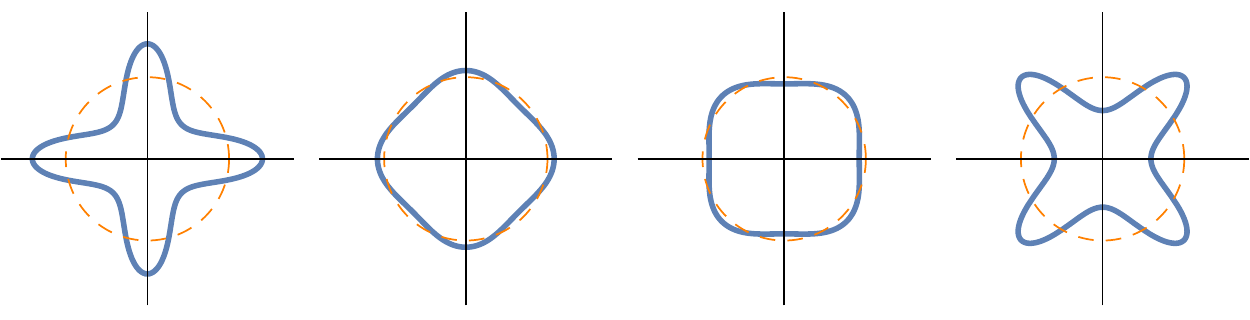} \\
\qquad $\alpha = -0.05$ \hfill $\alpha = -0.01$ \hfill $\alpha = 0.01$ \hfill $\alpha = 0.05$ \qquad\qquad \\[2mm]
\end{tabular}
\caption{Equilibrium shapes corresponding to $n=2$, $n=3$, $n=4$ and $\mu_1 = 0.3$, as obtained from \eqref{eq:change_xieta},\eqref{eq:ansatz_lineare} and \eqref{eq:linerised_xieta}, truncated to first order in $\alpha$, for increasing values of $\alpha$ (shown in the insets).}
\label{fig:linearised}
\end{figure}

We look for solutions of the equilibrium equations \eqref{eq:eq1_b}-\eqref{eq:eq3_b} as small perturbations of the circular solution
\begin{subequations}
\begin{align}
\theta(S) & = \frac{2\pi \, S}{L} + \alpha\, \theta_1(S), \qquad \xi(S) = \alpha L \, \xi_1(S), \\
\eta(S) & = -\frac{L}{2 \pi  (1+\mu_1)} + \alpha L \,\eta_1(S),
\end{align}
\label{eq:ansatz_lineare}
\end{subequations}
with $\alpha$ an arbitrary amplitude, $\theta_1(S)$, $\xi_1(S)$ and $\eta_1(S)$ unknown functions. After \modif{some simplification}, the linearised equilibrium equations (which define the operator $\Lop $) read
\begin{subequations}
\begin{align}
\theta_1'' - \frac{2\pi\mu_1 \mu_ 2}{L^2(1+\mu_1)} \xi_1 & = 0, \label{eq:lin_1}\\
\eta_1'(S) + \frac{2 \pi}{L}  \xi_1 & = 0, \label{eq:lin_2}\\
\xi_1'  + \frac{\theta'_1}{2\pi (1+\mu_1)}  - \frac{2\pi}{L}(1+\mu_1) \eta_1 & = 0.  \label{eq:lin_3}
\end{align}
\label{eq:linear_p}
\end{subequations}
We can write Eqs.\eqref{eq:linear_p} in the form of a system of first order ODEs $\by' = \bA \by + \bb$, with $\by=(\theta_1,\theta'_1,\eta_1,\xi_1)$, and look for solutions of the form $\by(S) = \mathrm{Re}(\by_0\, \er^{i \omega S})$. The corresponding linear problem has unique (trivial) solution, and no bifurcation occurs, except when $\det(i \omega \mathbf{I} - \bA) = 0$, where $\mathbf{I}$ is the identity matrix. Furthermore, continuity of the solution implies that $\omega = 2\pi n/L$, for some positive integer $n$. After some algebra, the bifurcation condition reads 
\begin{align}
4 \pi^2 (1+\mu_1)^2 (n^2-1-\mu_1)= \mu_1 \mu_2, 
\label{eq:bifurcation_condition}
\end{align}
where $n \in \mathbb{N}_+$. The mode $n=1$ is simply a translation of the circular solution. Hence, the first non-trivial mode is obtained by choosing $n=2$. For each value of $n$, there are two energetically equivalent eigenmodes, rotated by 45$^\circ$. We can study the deformations along only one of these eigenmodes by choosing the centre at the origin and the tangent $\theta(0)=0$. The corresponding 1-dimensional affine space is generated by
\begin{subequations}
\begin{align}
\theta_1^{(n)}(S) & = -\frac{2\pi}{n} \left(n^2 -\mu_1 -1\right)(1+\mu_1) \sin\Big(n\frac{2\pi \, S}{L}\Big), \\
\xi_1^{(n)}(S) & = -n\, \sin\Big(n\frac{2\pi \, S}{L}\Big), \\
\eta_1^{(n)}(S) & = \cos\Big(n\frac{2\pi \, S}{L}\Big).
\end{align}
\label{eq:linerised_xieta}
\end{subequations}
The amplitude $\alpha$ in \eqref{eq:ansatz_lineare} plays the role of an order parameter in our case. When $\alpha=0$, the solution is circular. A transition to a buckled shape is identified by $\alpha \neq 0$. To first order, the linearised solutions do not change the area, so that \eqref{eq:Area_xieta} still yields $\pi R_{\cc}^2$, to order $O(\alpha)$. In Fig.\ref{fig:linearised} we plot a gallery of shapes obtained by using the linearised solutions \eqref{eq:linerised_xieta}. \modif{The linear problem is non-trivial only at the critical point, when $\mu_1$ and $\mu_2$ are related by \eqref{eq:bifurcation_condition}, and the amplitude of the linear solution remains undetermined. Therefore, in order to fix the order parameter $\alpha$, we need to carry the analysis to higher orders. In our case, $\alpha$ is found by minimising the reduced energy function $g$ and, as expected, it will eventually depend on both material parameters $\mu_1$ and $\mu_2$.}

Due to their physical interpretation, it is natural to assume $\mu_1 \geq 0$ (i.e., a compressive force) and $\mu_2 > 0$. The value $\mu_2 = 0$ represents the non-physical situation where no energy is necessary to shorten or stretch the beam. The other limit, $\mu_2 \to + \infty$, \modif{leads to the inextensible case (discussed in detail below)} where $b \gg k/L^2$. From Eq.\eqref{eq:bifurcation_condition} we derive, at the critical point,
\modif{
\begin{equation}
\Big(\frac{\mu_1\mu_2}{4\pi^2}\Big)_{cr} = \big(1+(\mu_1)_{cr}\big)^2 \big(n^2-1-(\mu_1)_{cr}\big),
\label{eq:critical_mu}
\end{equation}}
so that the criticality condition is compatible with $\mu_2>0$ only when ($n>1$)
\begin{equation}
0 \leq (\mu_1)_{cr} < n^2-1. 
\end{equation}
This means that when $\mu_1 > n^2-1$, the mode $n$ cannot bifurcate and the circular solution is stable with respect to this and lower order perturbations. In particular, when $n=2$ the \modif{bifurcation point belongs to the interval} 
\begin{equation}
0 \leq (\mu_1)_{cr} < 3, \qquad (\mu_2)_{cr} > 0 . 
\end{equation}
To be more specific, we have that, at high compressions, if we take a value $3 < \mu_1 < 8$, mode $n=2$ cannot develop, but it is possible to observe a transition from the circular shape to a mode $n=3$. 

\modif{It is instructive to compare the critical pressure as given in \eqref{eq:critical_mu} with the one obtained for the inextensible limit. Tadjbakhsh and Odeh \cite{67Tadjbakhsh} find existence of buckled states for the ring when their applied dimensionless pressure $\hat{p}$ is larger than a critical value $\hat{p}_{cr} = 3$, with $n=2$. In their calculations they set the radius of the undeformed ring $R=1$ and the bending stiffness $k=1$. In order to compare our \eqref{eq:critical_mu} with their result, we now describe how to take the inextensible limit. In our notation, the inextensible limit corresponds to $\mu_2 \to +\infty$ and $\mu_1 \to 0$, such that the product $\mu_1\mu_2$ is constant. In fact $\mu_1$ and $\mu_2$ compare the external pressure and the bending modulus with the stretching modulus, $b$, which is what we want much larger than $pL/2\pi$ and $k/L^2$. At the same time, however, we have to balance $pL/2\pi$ and $k/L^2$, so that they are of the same order of magnitude and this implies $\mu_1 \mu_2=$constant. Comparing this with Tadjbakhsh and Odeh \cite{67Tadjbakhsh}, we find that their dimensionless pressure $\hat{p}$ corresponds to our figure $\mu_1\mu_2/4\pi^2 = pL^3/(8\pi^3 \, k)$, as they choose $k=1$ and $R = L/2\pi = 1$. If we now take Eq.\eqref{eq:critical_mu} and let $\mu_1 \to 0$, we readily find $\big(\mu_1\mu_2/4\pi^2\big)_{cr} = n^2-1$, which is equal to 3 for $n=2$, and coincides with the critical value given by  Tadjbakhsh and Odeh.}


In what follows we limit our analysis to the case $n=2$ and perform a full finite-dimensional reduction. This will allow us to derive a Landau-like potential which effectively describes the local behaviour of the elastic energy and, furthermore, provides a straightforward description of the number of equilibrium solutions, and their local stability, as the material parameters are varied.

\section{Finite-dimensional reduction (Landau expansion) for $n=2$}
\label{sec:Landaupotential}
There are two rather natural Banach spaces on which $W$ is well defined. These are $C^1([0,L])$, the continuously differentiable functions and the Sobolev space $H^1([0,L])$. However, the reduced energy derivation turns out to be very similar for the two spaces. For concreteness, we take $E = \big(C^1([0,L])\big)^3$, i.e., the space of triples $\bff = (f_1,f_2,f_3)$ of real continuously differentiable functions, with $L^2$-inner product
\begin{equation}
\langle \bff, \bg \rangle = \big\langle(f_1,f_2,f_3), (g_1,g_2,g_3)\big\rangle = \sum_{k=1}^{3}\int_{0}^{L}f_k(S) g_k(S) \dr S.
\end{equation}
The linear operator $\Lop$ is defined by \eqref{eq:linear_p}, and its kernel $N$ is found to be two-dimensional. As described in \S\ref{sec:linearanalysis} the two eigenvectors are energetically equivalent and allow generating all the rotated linear solutions. \modif{Without loss of generality, we take only the equilibrium shapes with horizontal tangent in $S=0$, i.e., we fix $\theta(0)=0$.} However, when projecting onto $\Np$, we need to include both generators in the definition of $Q$. Setting $n=2$ in \eqref{eq:linerised_xieta}, we find 
\begin{align}
\bv_1 & = \Big(-\pi (3-\mu_ 1) (\mu_1 + 1) \sin\left(\tfrac{4 \pi S}{L}\right),\
2\sin\left(\tfrac{4\pi S}{L}\right),\ \cos \left(\tfrac{4 \pi S}{L}\right) \Big),\\
\bv_2 & = \Big(\pi (3-\mu_1) (\mu_1+1) \cos \left(\tfrac{4 \pi  S}{L} \right),\
-2 \cos \left(\tfrac{4 \pi  S}{L}\right),\ \sin \left(\tfrac{4 \pi  S}{L}\right) \Big), 
\end{align}
and $N = \mathrm{span}(\bv_1,\bv_2)$. In this case, the projector $Q:E \to \Np$ is
\begin{align}
Q\bff & = \bff - \langle \bff,\bv_1\rangle\, \frac{\bv_1}{\|\bv_1 \|^2} - \langle \bff,\bv_2\rangle\, \frac{\bv_2}{\| \bv_2 \|^2} \notag \\
& =  (f_1,f_2,f_3) - \frac{1}{\| \bv_1 \|^2}\left[\sum_{k=1}^{3}\int_{0}^{L} f_k(S)v_{1,k}(S)\,\dr S \right] (v_{1,1},v_{1,2},v_{1,3})\notag \\
& - \frac{1}{\| \bv_2 \|^2}\left[\sum_{k=1}^{3}\int_{0}^{L} f_k(S)v_{2,k}(S)\,\dr S \right] (v_{2,1},v_{2,2},v_{2,3}).
\label{eq:Q_hypoarealis}
\end{align}
For fixed $\alpha$, higher order corrections belong to $\Np$ and are obtained by solving \eqref{eq:h}. Specifically, this is
$Q\nabla W(\alpha,\bw) = 0$, where $Q$ is given as in \eqref{eq:Q_hypoarealis} and $\nabla W$ is the triple of equilibrium equations \eqref{eq:eq_hypoarealis}, suitably rescaled so that they have the same dimensions. We then look for solutions of the form
\begin{align}
(L \theta,\xi,\eta) = (L \theta_{\cc},\xi_{\cc},\eta_{\cc}) + \alpha L \bv_1 + \alpha^2 L \bw_2(S) + \alpha^3 L \bw_3(S) + O(\alpha^3),
\label{eq:h_hypoarealis}
\end{align}
where $\bw_k = (\theta_k,\xi_k,\eta_k) \in \Np$. The equations for $\bw = \alpha^2 \bw_2+ \alpha^3 \bw_3$ turn out to be rather cumbersome, and are not reported for brevity. However, with the aid of Mathematica\textregistered \ they can be solved in terms of Fourier components. Specifically, we find
\begin{align}
\theta_2(S) & = \frac{\pi^2}{16} \left(\mu _1-3\right) \left(\mu _1+1\right)^2 \left(5 \mu _1-3\right) \sin \left(\tfrac{8 \pi S}{L}\right), \\
\xi_2(S) & = \frac{\pi}{2} \left(\mu _1-3\right) \left(\mu _1+1\right) \sin \left(\tfrac{8 \pi  S}{L}\right), \\
\eta_2(S) & = -\pi  \left(\mu_1-3\right) + \frac{5}{8} \pi  \left(\mu _1-3\right) \left(\mu _1+1\right) 
\cos \left(\tfrac{8 \pi  S}{L}\right), 
\end{align}
and
\begin{align}
\theta_3(S) & = b_{11} \sin \left(\tfrac{4 \pi S}{L}\right) + b_{12} \sin \left(\tfrac{12 \pi S}{L}\right), \label{eq:theta3}\\
\xi_3(S) & = b_{21} \sin \left(\tfrac{4 \pi S}{L}\right) 
+ b_{22} \sin \left(\tfrac{12 \pi S}{L}\right), \label{eq:xi3} \\
\eta_3(S) & = b_{31}\cos \left(\tfrac{4 \pi S}{L}\right) + b_{32} \cos \left(\tfrac{12 \pi S}{L}\right). \label{eq:eta3}
\end{align}
The coefficients $b_{11}$, $b_{12}$, $b_{21}$, $b_{22}$, $b_{31}$, $b_{32}$ and the explicit expressions for $(\theta,\xi,\eta)$ are reported in Appendix \ref{app:coefficienti}. It is worth noticing that, despite the fact that $(\theta_3,\xi_3,\eta_3)$ comprise $\cos(4 \pi S/L)$ terms, the vector $\bw_3$ belongs to $\Np$. In fact, it is possible to show, by direct computation, that the coefficients $b_{ij}$ are such that $\langle \bw_3,\bv_1\rangle = 0$ and $\langle \bw_3,\bv_2\rangle = 0$ (i.e, $Q\bw_3 = \bw_3$). 

Furthermore, the integral \eqref{eq:Area_xieta} yields the enclosed area, as a function of pressure and $\alpha$,
\begin{align}
A = \frac{L^2}{4 \pi  (\mu_1 + 1)^2} \left(1- 2 \pi^2 \alpha^2 (2 \mu_1^3-\mu_1^2+3)\right).
\label{eq:Area_O2}
\end{align}
The functions $(\theta,\xi,\eta)$, as given in \eqref{eq:h_hypoarealis}-\eqref{eq:eta3} and reported in Appendix \ref{app:coefficienti}, are then inserted into the elastic energy density. This is Taylor-expanded and integrated term-by-term to yield the reduced energy function $g(\alpha)$. The integrations are elementary, but rather cumbersome, so we again use a computer algebra software, such as Mathematica, to do the calculations. Therefore, the elastic energy \eqref{eq:W_final} is finally reduced to a much simpler \emph{function} of the order parameter $\alpha$, which correctly captures the nature of the critical points and the bifurcations. We find
\begin{align}
g(\alpha) & = a_2(\mu_1,\mu_2) \alpha^2 + a_4(\mu_1,\mu_2) \, \alpha ^4 + a_6 (\mu_1,\mu_2) \, \alpha ^6,
\label{eq:g_hypoarealis}
\end{align}
where 
\begin{equation}
a_2(\mu_1,\mu_2) = 2 \pi ^2 (3-\mu_1) \left(4 \pi^2 (1+\mu_1)^2 (3-\mu_1) - \mu _1 \mu _2\right),
\end{equation}
and $a_4$, $a_6$ are written in Appendix \ref{app:coefficienti_g}.	The coefficient $a_2(\mu_1,\mu_2)$ vanishes at the bifurcation, so that $g(\alpha)$ is not a Morse function at these points. When $a_2(\mu_1,\mu_2)=0$ and $a_4(\mu_1,\mu_2)>0$ the function $g$ is 4-determined and shows a supercritical pitchfork at the transition. By contrast, when $a_4(\mu_1,\mu_2)<0$, we observe a subcritical pitchfork bifurcation. The maximum degeneracy for $g(\alpha)$ occurs at particular values of $\mu_1$ and $\mu_2$ such that $a_2(\mu_1,\mu_2)=0$ and $a_4(\mu_1,\mu_2)=0$. Since we only have two parameters and two polynomial equations, we can generically have only a finite number of solutions. In fact, if we calculate the polynomial resultant of $a_2$ and $a_4$ to eliminate $\mu_2$, we find that $a_2=a_4=0$ is equivalent to 
\begin{equation}
11 \mu_1^2 - 90 \mu_1 +27 = 0,
\end{equation}
so that the \emph{only acceptable point} of the parameter space $(\mu_1,\mu_2)$ where $a_2=a_4=0$ is 
\begin{equation}
(\mu_1,\mu_2) \approx (0.312,585).
\label{eq:tricritical_values}
\end{equation} 
However, when we choose these values for the parameters, we get $a_6 >0$, so that the function $g(\alpha)$ is \emph{at most 6-determined}. This means that the local bifurcations around the circular solutions are correctly described considering the truncation of the elastic energy at the $6^{th}$-degree term. Furthermore, it is possible to check that the expansion of \eqref{eq:h_hypoarealis} to order $O(\alpha^3)$ it is sufficient to determine exactly the coefficients of $g$ up to its $6^{th}$-degree. Hence, $g(\alpha)$ as given in \eqref{eq:g_hypoarealis} rigorously reproduces the nature of the bifurcations and the stability of the solutions.

\section{Analysis of the phase transitions}
\label{sec:phasetransitions}
In Catastrophe Theory parlance, $g$ is a symmetric butterfly catastrophe, and the organizing centre of the singularity is $+ \alpha^6$ \cite{78Golubitsky,Demazure}. Its normal form is
\begin{equation}
g(\alpha) = \alpha^6 + c_4 \alpha^4 + c_2 \alpha^2,
\end{equation}
where we have simply redefined the control parameters as $c_2 = a_2/a_6$ and $c_4 = a_4/a_6$. The reduced function is even, and this is a consequence of the symmetry of the original energy functional. If we wish to include the effect of possible imperfections we should consider a universal unfolding of the butterfly singularity. In this case, the reduced function also comprises odd powers of $\alpha$. However, we consider here only the symmetric case.

\begin{figure}
\begin{center}
\includegraphics[width=0.43\textwidth]{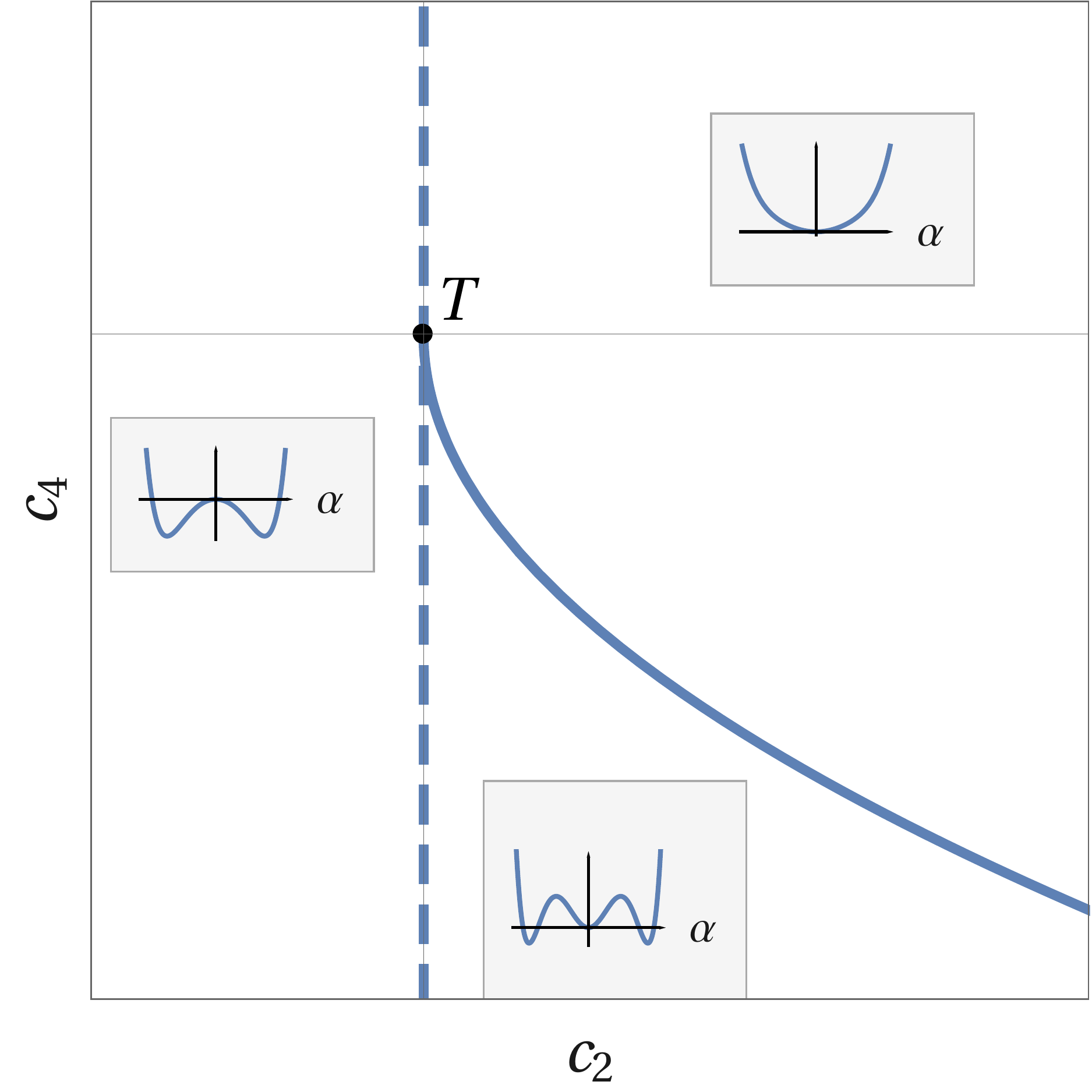} \hfill 
\includegraphics[width=0.45\textwidth]{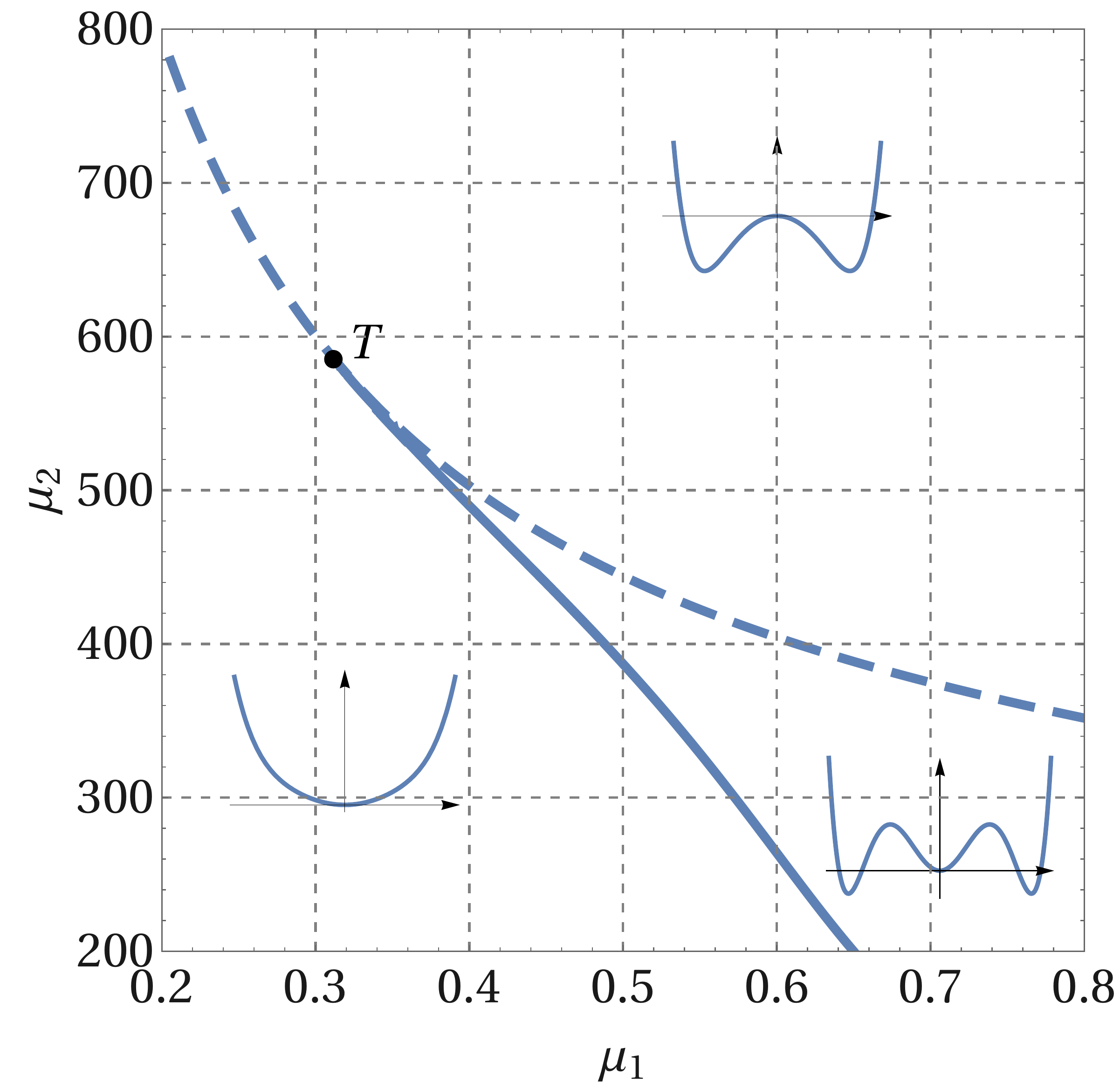}
\end{center}
\caption{\modif{Bifurcation set in the mathematical control plane $(c_2,c_4)$ (left) and in the physical parameter plane $(\mu_1,\mu_2)$ (right). The reduced potential has the same qualitative features within each region, as shown in the insets. The number of critical points changes, and thus a transition occurs, when crossing the boundary lines. Point ``T'' corresponds to the tricritical point $(c_1,c_2) = (0,0)$, i.e., $(\mu_1,\mu_2) \approx (0.312,585)$.}}
\label{fig:bifurcationSet}
\end{figure}

The critical manifold, defined by $\D{g}{\alpha}=0$, projects onto the control plane $(c_2,c_4)$ and provides the bifurcation set shown in Fig.\ref{fig:bifurcationSet}, which decomposes the control plane into homogeneous regions. Any potential within a region shows the same qualitative features and the same number of critical points. The number and nature of critical points change when crossing the boundaries of these region. 
The special point ``T'' in Fig.\ref{fig:bifurcationSet} has coordinates $(c_2,c_4) = (0,0)$, and corresponds to the maximum degeneracy, where the values of $\mu_1$ and $\mu_2$ are those given in \eqref{eq:tricritical_values}.

\begin{figure}
	\begin{center}
\includegraphics[width=0.9\textwidth]{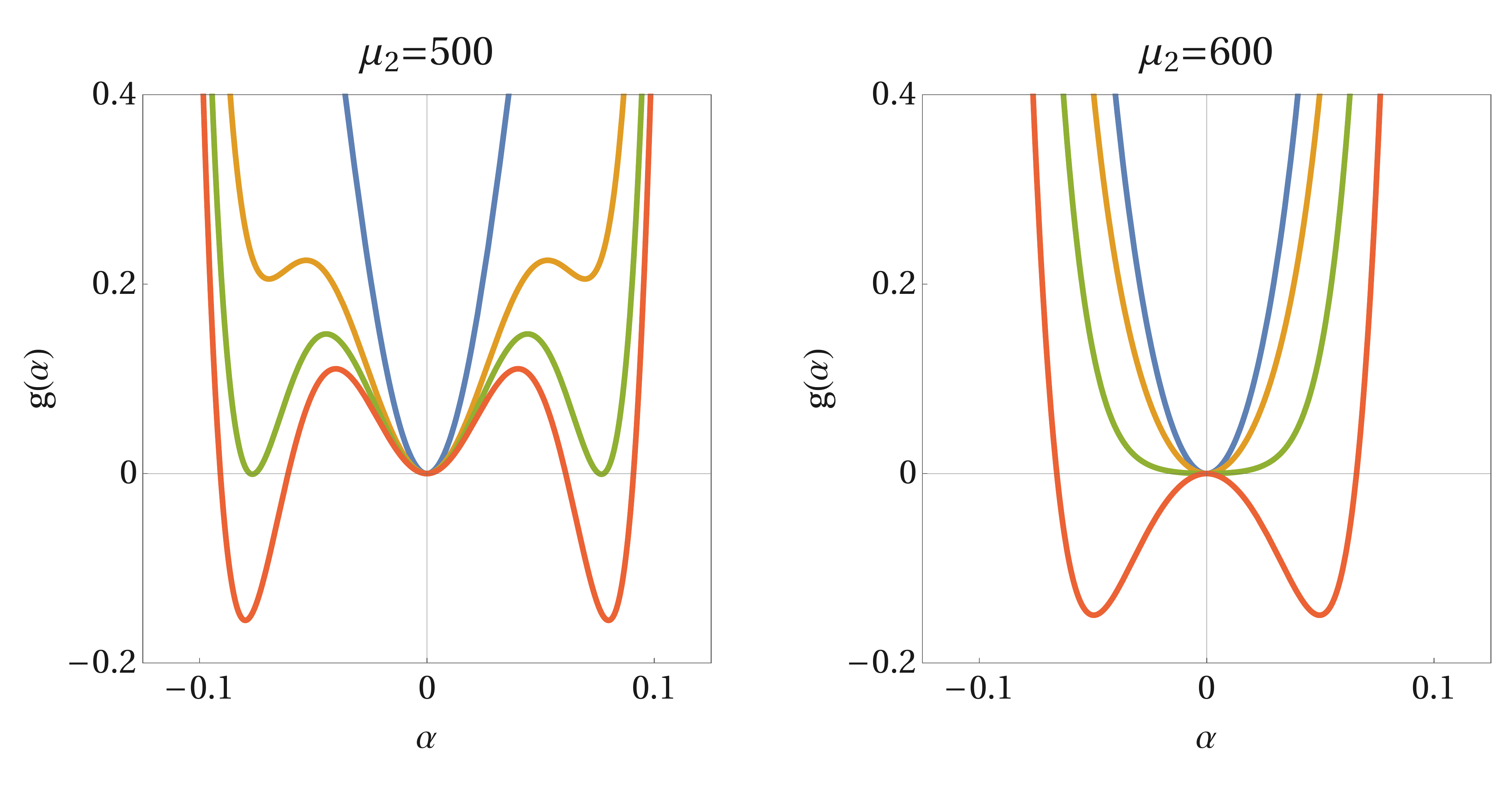}
	\end{center}
	\caption{Reduced energy profiles for increasing values of the dimensionless pressure $\mu_1$: first-order transition (left, $\mu_2 = 500$), and second order transition (right, $\mu_2 =600$). }
	\label{fig:transitions}
\end{figure}

In physical terms, the dashed line identifies a second-order transition, where the system undergoes a continuous transformation from circular to buckled shaped. By contrast the solid line marks a first-order transition where the deformation is discontinuous. Examples of $g(\alpha)$ for various values of the material parameters, corresponding to second and first-order transitions, are shown in Fig. \ref{fig:transitions}. Point ``T'' is a \emph{tricritical point} where a phase transition of a complex kind takes place involving the meeting of a second-order transition with a line of first-order transitions.
We observe that for $\mu_2$ greater than the tricritical value $585$, i.e., in the nearly inextensible case, the transition is second-order. By contrast, in ``soft materials'', with $\mu_2 < 585$, we obtain a discontinuous snap-through buckling (first-order transition) with increasing pressure.
This is apparent also from Fig.\ref{fig:Aversusp} where we analyse how the area depend on pressure (or vice-versa, if we imagine to impose an area constraint).

{\color{blue}
\begin{figure}
	\begin{center}
		\includegraphics[width=0.6\textwidth]{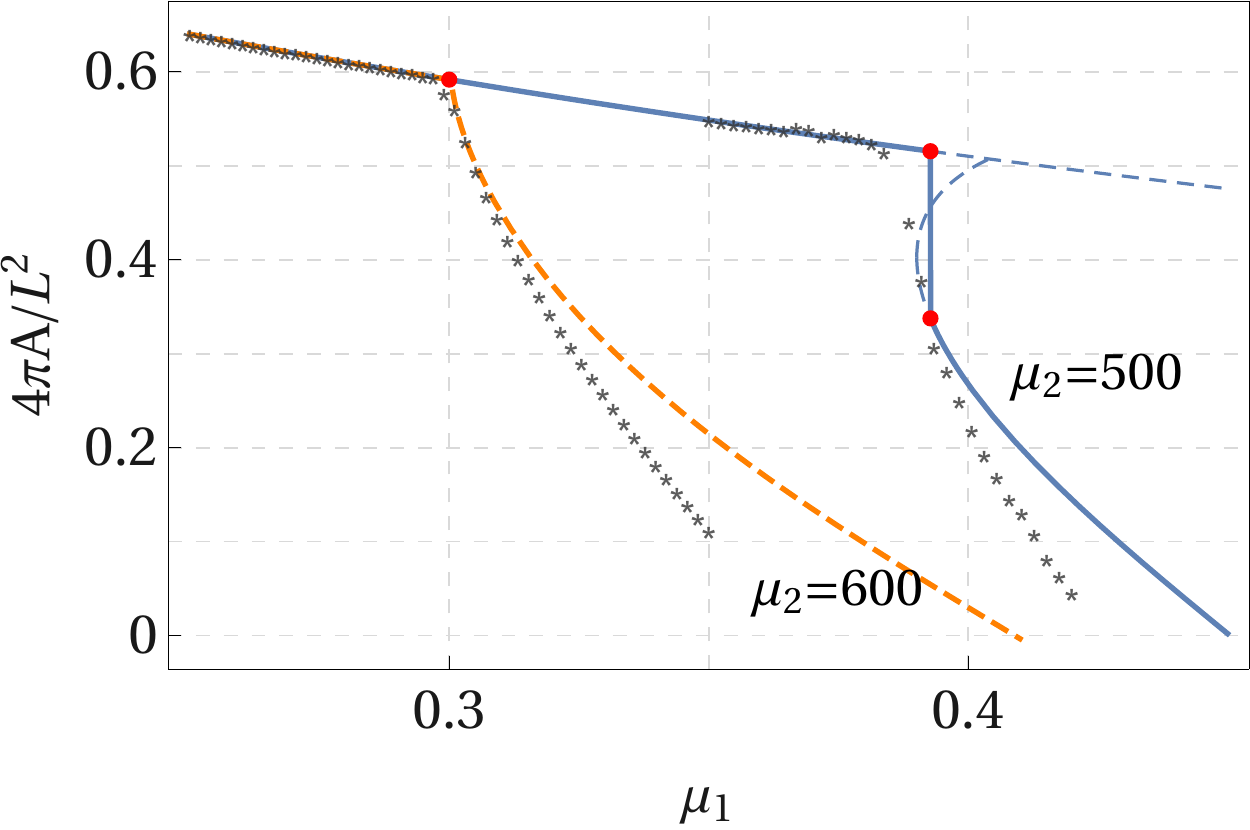}
	\end{center}
	\caption{Enclosed area against pressure, as given in \eqref{eq:Area_O2}, for two values of $\mu_2$. The order parameter $\alpha$ is obtained by minimizing the reduced energy $g(\alpha)$, as given in \eqref{eq:g_hypoarealis}. Initial straight lines correspond to compressed circular solutions, with $\alpha=0$. Red dots mark the transitions to buckled configurations, where $\alpha \neq 0$. A Maxwell criterion is adopted for the first-order transition with $\mu_2=500$. \modif{For comparison, grey stars show the numerical results, obtained by integrating the equilibrium equations \eqref{eq:eq1}-\eqref{eq:eq4}.}}
	\label{fig:Aversusp}
\end{figure}
}
	
In the figure, the transition with $\mu_2 = 500$ has been calculated using a Maxwell criterion: the buckling occurs when $g(\alpha)$ has (at least) two distinct relative minima with the same minimizing value. It is worth noticing that a first-order transition implies the existence of metastable states, whose characteristic signature are bistability effects and hysteresis, as observed in some problems of cell mechanics \cite{13Banerjee,19Giomi}.

\section{Conclusions}
\label{sec:conclusions}
In this article we have reviewed some fundamental mathematical techniques that we believe are very helpful to study bifurcation problems. In elasticity  similar methods are routinely used (for example Lyapunov-Schmidt reduction). However, these do not take full advantage of the original variational structure, so that the analysis of the equilibrium solutions, bifurcation conditions and local stability are treated as separate problems. Most of the times they require to check the positive definiteness of the second variation. A problem that is usually intractable from an analytical standpoint (although clever geometrical methods can simplify the analysis in some cases \cite{87Maddocks,17Goriely}).

We have shown that Singularity theory allows us to simplify the original functional to a reduced function $g$ that provides an intuitive but rigorous account of the full bifurcation scenario, including the local stability of the solutions. The determinacy of $g$ gives a bound for its Taylor expansion, so that it is guaranteed that a truncation of the reduced function at the correct order is sufficient to capture all the qualitative information about the local unfolding of the energy functional. 

These techniques are then applied to study the shape bifurcations of an extensible two-dimensional elastic ring subject to uniform pressure. The origins of this problem trace back to more than a century ago, when the scope was to investigated the collapse of buried pipelines under pressure. However, the subject has received renewed interest by a heterogeneous community of physicists, applied mathematicians and engineers and is now a powerful prototype problem with potential applications to various softly constrained physical systems. Applications to biological tissues and cells seems to be the most promising. We find that, depending on the stretching modulus, the transition can either be second or first-order. In particular, first-order transition implies the existence of bistability effects and hysteresis, and these have been observed in some problems of cell mechanics.

Despite the wide theoretical applicability, the main practical limitation of the present approach is the solution of Eq.\eqref{eq:h}. Usually, this is feasible only for one, or maybe two, order parameters. It is worth mentioning that symmetry arguments could be directly included in the analysis to possibly simplify the computations. Finally, this reduction can be adapted to address also simple free-boundary problems or unilateral constraints.

\section*{Acknowledgments}
I wish to thank P. Nardinocchi, M. Curatolo and G. Napoli for instructive conversations about several key aspects of the present problem. This work was supported by the Italian Ministry of University and Research through the grant N.2017KL4EF3 ``Mathematics of active materials: from mechanobiology to smart devices''.

\appendix
\section{Proof of the splitting lemma \ref{thm:splitting}}
\label{app:proofSL}
The proof that we report here mostly follows \cite{84Geymonat}. For simplicity, let us write $F=Q\nabla W: U\subset E \to \Np$, where $Q$ is the orthogonal projector onto $\Np$. Equation \eqref{eq:h} is then $F(x,y) = 0$, where we have used the notation $(x,y)$ for point $x+y$ in $N \oplus \Np \simeq E$. 

For a given $x \in N$ the partial derivative $D_2 F(x,\cdot)$ in $y=0$ can be calculated from the following expansion
\begin{align}
F(x,\eps y) & = Q \nabla W(x,\eps y) 
= Q\nabla W(x,0) + \eps Q (D\nabla W(x,0))[y] + o(\eps).
\end{align}
Evaluating this expression in $x=0$, and using \eqref{eq:L_and_gradW}, we derive
\begin{equation}
D_2 F(0,0) = Q\Lop .
\end{equation}
By assumption, $Q\Lop $ restricted to $\Np$ is an isomorphism of $\Np$ to itself. Thus, the implicit function theorem guarantees that the equation $F(x,y)=0$, i.e., equation \eqref{eq:h}, uniquely defines a function $y=h(x)$ near $(0, 0)$, and $h(0)= 0$. Furthermore, since $x \in N = \ker \Lop $, 
\begin{align}
F(\eps x,0) & = Q \nabla W(\eps x,0) 
= Q\nabla W(0,0) + \eps Q (D\nabla W(0,0))[x] + o(\eps) \notag \\
& = \eps Q\Lop x + o(\eps) = o(\eps).
\end{align}
Thus, $D_1 F(0,0) = 0$ so that $h'(0)= 0$. \\

Consider now the function $f:\Np \to \Np$ 
\begin{equation}
f(z) = W(x, h(x) + z) - W(x , h(x)),
\end{equation}
where $x$ is taken as a parameter and $z \in \Np$. We can expand $f$ to read its first and second derivative
\begin{align}
f(\eps z) & = f(0) + \eps f'(0) [z] + \frac{1}{2}\eps^2 f''(0)[z,z] + o(\eps^2) \notag \\
& = \eps DW(x, h(x))[z] + \frac{1}{2}\eps^2 D^2 W(x, h(x))[z,z] + o(\eps^2).
\end{align}
Evaluating this expression in $x=0$ (taking $x$ as a parameter), we obtain
\begin{equation}
f'(0)|_{x=0} = 0, \qquad f''(0)|_{x=0} = \Lop  .
\end{equation}
Since $\Lop $ is non-degenerate if restricted to $\Np$, it is possible to apply the parametric version of Morse lemma \cite{69Palais,84Geymonat} to $f(z)$. Then, there exists an origin preserving $k$-diffeomorphism $z \mapsto r(x,z)$ such that
\begin{align}
W(x, h(x) + r(x,z)) - W(x, h(x)) = \frac{1}{2} \langle \Lop z, z \rangle
\end{align}
Therefore, with the change of variable $\Phi: (x,y) \mapsto (x,h(x) + r(x,y))$ we have, in a neighbourhood of $(0,0)$,
\begin{align}
W\circ\Phi(x,y) = \frac{1}{2} \langle \Lop y, y \rangle + W(x + h(x)).
\end{align}
This concludes the proof.

\section{Coefficients}
\label{app:coefficienti}
We report here the long coefficients for equations \eqref{eq:theta3}-\eqref{eq:eta3}.
\begin{align}
b_{11} & =  -\frac{\pi ^3 \left(\mu _1-3\right)^2 \left(\mu _1+1\right)^2}
{4 \left(-\mu _1+24 \pi +3\right) \left(\pi ^2 \left(\mu _1-3\right)^2 \left(\mu _1+1\right)^2+5\right)} \times \notag \\
& \left[\mu _1 \left(-6 \mu _1^2+\mu _1-90\right)+4 \pi \left(\mu _1 \left(80 \mu _1-93\right)+99\right)+63\right],\\
b_{12} & = \frac{7}{768} \pi ^3 \left(\mu _1-3\right)^2 \left(\mu _1+1\right)^3 \left(13 \mu _1-3\right),
\end{align}
\begin{align}
b_{21} & =  -\frac{\pi ^2 \left(\mu _1-3\right)^2 \left(\mu _1+1\right)}{4 \left(-\mu _1+24 \pi +3\right) 
	\left(\pi ^2 \left(\mu _1-3\right)^2 \left(\mu _1+1\right)^2+5\right)} \times \notag \\
	& \big[-8 \pi ^3 \left(\mu _1-3\right) \left(\mu _1 \left(20 \mu_1-33\right)+27\right)(\mu _1+1)^2 \notag \\
	&+\pi ^2 \left(\mu_1-3\right) \left(\mu _1 \left(\mu _1 \left(3 \mu_1-1\right)+33\right)-27\right) \left(\mu _1+1\right)^2 \notag \\
	& +3 \left(\mu_1+1 \right)^2 + 32 \pi  \left(3-5 \mu _1\right)\big],\\
b_{22} & = \frac{7}{128} \pi ^2 \left(\mu _1-3\right) \left(\mu_1+1\right)^2 \left(7 \mu _1-9\right),
\end{align}
\begin{align}
b_{31} & = \frac{\pi^2 \left(\mu _1-3\right)^2\left(\mu _1+1\right)}
{4 \left(-\mu _1+24 \pi +3\right) \left(\pi ^2 \left(\mu _1-3\right)^2 \left(\mu _1+1\right)^2+5\right)} \times \notag \\
& \big[12 \pi ^3 \left(\mu _1-3\right) \left(13 \mu_1-3\right) \left(\mu_1+1\right)^2 \notag \\
& -\pi ^2 \left(\mu _1-3\right) 
\left(\mu _1 \left(\mu _1+24\right)-9\right) \left(\mu _1+1\right)^2
+6 \left(\mu _1+1\right)^2+64 \pi  \left(3-5 \mu _1\right)\big], \\
b_{32} & = \frac{13}{256} \pi^2 \left(\mu_1+1\right)^2 \left(7 \mu _1-9\right) \left(\mu _1-3\right).
\end{align}

Explicitly, we find
\begin{align}
\theta(S) & = \frac{2 \pi  S}{L} - \pi  \alpha  \left(3-\mu_1\right) \left(\mu _1+1\right) \sin \left(\tfrac{4 \pi  S}{L}\right) \notag \\
& + \frac{\pi^2}{16} \alpha^2\left(\mu _1-3\right) \left(\mu _1+1\right)^2 \left(5 \mu _1-3\right) \sin \left(\tfrac{8 \pi S}{L}\right) \notag \\
& +  \alpha^3 \Big[b_{11}\sin \left(\tfrac{4 \pi  S}{L}\right) + b_{12} \sin \left(\tfrac{12 \pi S}{L}\right)\Big].\\
\xi(S) & = 2 L \alpha  \sin \left(\tfrac{4 \pi  S}{L}\right) 
+ \frac{\pi}{2}  L\,\alpha^2 \left(\mu _1-3\right) \left(\mu _1+1\right) \sin \left(\tfrac{8 \pi  S}{L}\right) \notag \\
& + L \alpha^3 \Big[b_{21} \sin \left(\tfrac{4 \pi S}{L}\right) + b_{22} \sin \left(\tfrac{12 \pi S}{L}\right) \Big].\\
\eta(S) & = -\frac{L}{2 \pi  (\mu_1+1)} + L \alpha  \cos \left(\tfrac{4 \pi  S}{L}\right) \notag \\
& + L \alpha^2 \left(\frac{5}{8} \pi  \left(\mu _1-3\right) \left(\mu _1+1\right) 
\cos \left(\tfrac{8 \pi  S}{L}\right)-\pi  \left(\mu _1-3\right)\right) \notag \\
& + L \alpha^3 \Big[ b_{31} \cos \left(\tfrac{4 \pi  S}{L}\right) + b_{32}\cos \left(\tfrac{12 \pi  S}{L}\right)\Big].
\end{align}

\section{Coefficients of the reduced energy function}
\label{app:coefficienti_g}

\begin{align}
a_4(\mu_1,\mu_2) & = -\frac{\pi ^4 \left(\mu _1-3\right) \left(\mu _1+1\right)}
{32 \left(-\mu _1+24 \pi +3\right) \left(\pi ^2 \left(\mu _1^2-2 \mu_1-3\right)^2+5\right)} \times \notag \\ 
&	\Big[4 \pi ^4 \left(3-5 \mu _1\right)^2 \left(\mu_1-3\right)^4 \left(\mu _1+1\right)^5
-96 \pi ^5 \left(3-5 \mu _1\right)^2 \left(\mu _1-3\right)^3 \left(\mu_1+1\right)^5 \notag \\
&	-8 \pi ^3 \left(\mu _1-3\right) 
\big(75 \mu _2 \mu_1^5-372 \mu _2 \mu _1^4+2 \left(729 \mu _2-1810\right)\mu_1^3 \notag \\
&	+\left(21012-3348 \mu _2\right) \mu_1^2+9 \left(99 \mu _2-2828\right) \mu _1+19548\big) \left(\mu_1+1\right)^2 \notag \\
&	+8 \pi  \mu_1 \left(905 \mu _1^3-4593 \mu_1^2+963 \mu _1-3267\right) \mu_2 \notag \\
&	-\mu_1 \left(67 \mu _1^4+12\mu _1^3+546 \mu _1^2-5076 \mu _1+4563\right) \mu _2 \notag \\
&	+\pi^2 \left(\mu _1^2-2 \mu _1-3\right)^2 \big(25 \mu _2 \mu_1^5-124 \mu _2 \mu _1^4+\left(486 \mu _2-268\right) \mu_1^3 \notag \\
&	-4 \left(279 \mu _2-7\right) \mu_1^2+3 \left(99 \mu_2-3980\right) \mu_1+8244\big)\Big],
\end{align}

\begin{align}
a_6 & (\mu_1, \mu_2) = \frac{\pi ^6 \left(\mu _1-3\right)^2 \left(\mu _1+1\right)^2}
{32768 \left(\mu _1-24 \pi -3\right)^2 \left(\pi ^2 \left(\mu _1^2-2 \mu _1-3\right)^2+5\right)^2} \times \notag \\
&	\Big[-9408 \pi ^7 \left(3-13 \mu _1\right)^2 \left(\mu _1-3\right)^7 \left(\mu _1+1\right)^8 \notag \\
&  +112896 \pi ^8 \left(3-13 \mu _1\right)^2 \left(\mu _1-3\right)^6 \left(\mu _1+1\right)^8 \notag \\
&	+\left(\mu _1-3\right)^2 \mu _1 \big(354481 \mu_1^5+756721 \mu _1^4+4947034 \mu _1^3+24449106 \mu _1^2   \notag \\
&  -26947611 \mu_1+5576877\big) \mu_2 - 16 \pi  \mu _1 \big(1604115 \mu_1^6-2550374 \mu _1^5-23872107 \mu _1^4 \notag \\
& +111841740 \mu_1^3-254712627 \mu _1^2	+235371258 \mu _1-49445397\big) \mu _2 \notag \\
& -48 \pi ^5 \left(\mu_1^2-2 \mu _1-3\right)^4 	\big(8281 \mu _2 \mu _1^7-64370 \mu _2 \mu _1^6+\left(381343 \mu _2+331240\right) \mu _1^5 \notag \\
& -4 \left(119943 \mu_2+121030\right) \mu _1^4+\left(1865799 \mu _2-1485680\right) \mu_1^3 \notag \\
& -18 \left(73377 \mu _2+13720\right) \mu _1^2 +27 \left(2739 \mu_2+13720\right) \mu _1-52920\big) \notag \\
& -32 \pi ^3 \left(\mu _1^2-2 \mu_1-3\right)^2 
\big(124215 \mu _2 \mu _1^7-312238 \mu _2 \mu _1^6-9 \left(170487 \mu _2-356470\right) \mu_1^5 \notag \\
&	+2 \left(7599150 \mu_2-2214349\right) \mu_1^4-9 \left(2533599 \mu _2-2970428\right) \mu_1^3 \notag \\
&	+18\left(1332369 \mu _2-3125498\right) \mu_1^2+\left(61882974-8221905 \mu _2\right) \mu_1-25745202\big) \notag \\
& +4 \pi^6 \left(\mu _1^2-2 \mu _1-3\right)^4 \big(8281 \mu_1^{10}-70070 \mu _1^9+130389 \mu_1^8+281848 \mu_1^7 \notag \\
& +98 \left(12168 \mu_2-7759\right) \mu_1^6-4 \left(3405436 \mu _2+173313\right) \mu_1^5 \notag \\
& +18 \left(8319824 \mu _2+2723273\right) \mu_1^4-72 \left(3984828 \mu_2-1037575\right) \mu _1^3 \notag \\
& +9 \left(17087184 \mu _2+678013\right) \mu _1^2-54 \left(286920 \mu _2+317569\right) \mu_1+2575881\big) \notag \\
&	+2 \pi ^2 \big(41405 \mu _2 \mu_1^{12}-476517 \mu_2 \mu_1^{11} + \left(2499091 \mu _2+708962\right) \mu_1^{10} \notag \\ 
& -\left(6424643 \mu_2+4781452\right) \mu_1^9+\left(15178394-6443198 \mu _2\right) \mu_1^8 \notag \\ 
& + 6 \left(15084045 \mu _2-4210328\right) \mu_1^7+\left(95619590 \mu_2+43802596\right) \mu _1^6 \notag \\
& - 2 \left(496619875 \mu_2+138196068\right) \mu_1^5+567 \left(3615575 \mu_2+569724\right) \mu_1^4 \notag \\ 
& + \left(861527664-6024009177 \mu _2\right) \mu _1^3 - 81 \left(19657465 \mu _2+7144038\right) \mu_1^2 \notag \\ 
& + 1701 \left(572357 \mu_2-268444\right) \mu _1+294412482\big) \notag \\
&	+\pi ^4 \left(\mu _1^2-2 \mu_1-3\right)^2 \big(8281 \mu _2 \mu _1^{12}-122337 \mu_2 \mu_1^{11}+\left(914743 \mu _2+331240\right) \mu_1^{10}\notag \\
& -35 \left(104109 \mu_2+80080\right) \mu_1^9 +\left(7655386 \mu _2+5215560\right) \mu_1^8 \notag \\
& - 2 \left(5400261 \mu _2-5636960\right) \mu_1^7 +\left(58481838 \mu_2-30415280\right) \mu_1^6 \notag \\ 
&  -2 \left(88161623 \mu _2+13865040\right) \mu_1^5 + 5 \left(539280153 \mu_2+441504272\right) \mu_1^4 \notag \\ 
& -3 \left(2306970711 \mu _2+1037986240\right) \mu_1^3 +9 \left(188236755 \mu _2+719716744\right) \mu_1^2 \notag \\
& + 81 \left(5300277 \mu _2-61790576\right) \mu _1 + 2596103784\big)\Big].
\end{align}

\bibliographystyle{unsrt}
\bibliography{ReferencesEH}

\end{document}